\documentclass[namedate,webpdf,traditional,medium]{oup-authoring-template}
\usepackage{hyperref}

\onecolumn 

\usepackage{subfigure}
\graphicspath{{Figures/}}
\usepackage{epsfig}

\theoremstyle{thmstyleone}%
\theoremstyle{thmstyletwo}%
\theoremstyle{thmstylethree}%

\usepackage{amsmath}

\begin{document}

\journaltitle{}
\copyrightyear{2022}

\firstpage{1}

\title[Multivariate longitudinal modeling with continuous time-varying endogenous covariates]{Multivariate longitudinal modeling of cross-sectional and lagged associations between a continuous time-varying endogenous covariate and a non-Gaussian outcome}

\author[1]{Chiara Degan}
\author[1]{Bart J.A. Mertens}
\author[1]{Jelle Goeman}
\author[2]{Nadine A. Ikelaar}
\author[2]{Erik H. Niks}
\author[3]{Pietro Spitali}
\author[1]{Roula Tsonaka}

\authormark{Degan et al.}

\address[1]{\orgdiv{Department of Biomedical Data Sciences}, \orgname{Leiden University Medical Center}, \country{The Netherlands}}
\address[2]{\orgdiv{Department of Neurology}, \orgname{Leiden University Medical Center}, \country{The Netherlands}}
\address[3]{\orgdiv{Department of Human Genetics}, \orgname{Leiden University Medical Center}, \country{The Netherlands}}

\abstract{In longitudinal studies, time-varying covariates are often endogenous, meaning their values depend on both their own history and that of the outcome variable. This violates key assumptions of Generalized Linear Mixed Effects Models (GLMMs), leading to biased and inconsistent estimates. Additionally, missing data and non-concurrent measurements between covariates and outcomes further complicate analysis, especially in rare or degenerative diseases where data is limited. To address these challenges, we propose an alternative use of two well-known multivariate models, each assuming a different form of the association. One induces the association by jointly modeling the random effects, called Joint Mixed Model (JMM); the other quantifies the association using a scaling factor, called Joint Scaled Model (JSM). We extend these models to accommodate continuous endogenous covariates and a wide range of longitudinal outcome types. A limitation in both cases is that the interpretation of the association is neither straightforward nor easy to communicate to scientists. Hence, we have numerically derived an association coefficient that measures the marginal relation between the outcome and the endogenous covariate. The proposed method provides interpretable, population-level estimates of cross-sectional associations (capturing relationships between covariates and outcomes measured at the same time point) and lagged associations (quantifying how past covariate values influence future outcomes), enabling clearer clinical insights. We fitted the JMM and JSM using a flexible Bayesian estimation approach, known as Integrated Nested Laplace Approximation (INLA), to overcome computation burden problems. These models will be presented along with the results of a simulation study and a natural history study on patients with Duchenne Muscular Dystrophy.}

\maketitle

\section{Introduction}\label{intro}
In longitudinal studies, repeated measurements are collected from different subjects over time, leading to correlated observations within individuals. To appropriately model this correlation structure, Linear Mixed Effects Models (LMMs) and Generalized Linear Mixed Effects Models (GLMMs) \citep{diggle_analysis_2002, robert_e_weiss_modeling_2005, molenberghs_models_2005, wu_mixed_2009} are commonly used, allowing inferences at both the population and the subject levels. 

However, implementing these models can be challenging when time-varying covariates are involved, especially when those covariates are endogenous. In longitudinal settings, time-varying independent variables can be classified as either exogenous or endogenous. Exogenous variables, such as chronological age or daily climate change, vary over time based solely on their previous values, independent of the outcome \citep{diggle_analysis_2002, qian_linear_2020}. For instance, in a study assessing the effect of air pollution on asthma, the amount of air pollution is an exogenous variable because it changes independently of the asthma status of the subject. Endogenous variables, on the other hand, are influenced by both their own history and past values of the response variable \citep{diggle_analysis_2002, qian_linear_2020}. For example, in evaluating the association between asthma severity and inhaler use, the treatment variable (inhaler dosage) is an endogenous variable because it may adapt in response to symptoms progression. Another example of endogenous variables can be taken from our motivating study on Duchenne Muscular Dystrophy \citep[DMD,][]{duan_duchenne_2021}, a rare genetic neuromuscular disease characterized by progressive muscle wasting. The study follows patients over time, collecting repeated measurements of physical tests and protein biomarkers, with the aim of identifying biological indicators of disease progression. In this case, the proteins are also endogenous covariates: their levels may reflect or respond to the stage of disease progression, which is captured through physical tests. 
Because of the relationship that exists between endogenous and dependent variables, standard mixed effect models may no longer be appropriate to assess the association between the variables. Similar to the response variable, the time-varying covariate varies over time, and its stochastic evolution (e.g. its measurement error) should also be studied. The covariate process must be specified and cannot be ignored, otherwise estimates of the parameters of interest may be biased and inconsistent \citep{qian_linear_2020}, which can lead to wrong conclusions. 

A second challenge arises when the longitudinal response and the covariate are collected at different points in time, and one variable may have missing data while the other has observed values. Conventional applications of LMMs and GLMMs exclude all measurements of visits in which at least one variable of interest is missing. By default, the most used R-packages apply (i.e., \href{https://cran.r-project.org/web/packages/lme4/index.html}{lme4}) or need (i.e.,\href{https://cran.r-project.org/web/packages/nlme/index.html}{nlme}) listwise deletion methods \citep{buuren_flexible_2018}. Consequently, significant bias is introduced in the estimates, and valuable information is removed, reducing statistical power \citep{little_statistical_2019}. This is a crucial aspect, especially in biological or medical settings, where it is common to have data with a small number of patients, such as those from rare diseases, or data with short follow-up, such as those from degenerative diseases. Furthermore, the complexity of clinical assessments, which typically require collecting a wide range of physiological and functional parameters, inherently increases the likelihood of missing data. It is even more pronounced in pediatric studies, where obtaining consistent measurements is often more difficult due to limited patient cooperation. Again, this is exemplified in our DMD study, where relying only on complete records would discard nearly half of the available data.
Beyond the limitations of listwise deletion, missing data introduces additional complexity when modeling temporal relationships between covariates and outcomes. In longitudinal studies, it is often necessary to estimate lagged effects, that is, how a covariate measured at one time point influences the outcome at a later time. However, the presence of missing values in either the covariate or the outcome limits the ability to estimate potential lagged effects within a GLMM framework \citep{diggle_analysis_2002}. Importantly, estimating only the immediate effect (same time point) of a covariate on the outcome is often insufficient, and may fail to reflect the real-world mechanism generating the data. Several examples in the literature demonstrated that ignoring the lagged effect can lead to incorrect results, both in biology settings \citep{fisher_time-dependent_1999, rizopoulos_joint_2012} and others \citep{li_considering_2023}. 
Therefore, while using GLMMs, we have to be mindful that some assumptions may not necessarily be consistent with the nature of the data, and if we are not careful, we risk losing a large amount of data or obtaining less accurate estimates. 

Multivariate models, on the other hand, can be used to model the two processes jointly and overcome the issues just described. They accommodate endogenous process variability and utilize all available data. In this paper we focus on two multivariate models: the Joint Mixed Model \citep[JMM,][]{robert_e_weiss_modeling_2005, fitzmaurice_longitudinal_2008, verbeke_analysis_2012, mcculloch_joint_2016} and the Joint Scaled Model \citep[JSM,][]{rizopoulos_joint_2012}. The former, models the association via the variance-covariance matrix of the correlated random effects; the latter, instead, induces the association by copying and scaling the linear predictor of the endogenous covariate into the linear predictor of the outcome. The JMM has been used for normal response variables \citep{drikvandi_framework_2024} and was extended to the case of normal-binary responses \citep{fitzmaurice_longitudinal_2008, iddi_joint_2012, amini_longitudinal_2018, delporte_joint_2022}, normal-ordinal responses \citep{ivanova_mixed_2016, delporte_joint_2025} and normal-count responses \citep{kassahun_joint_2013}. More complicated combinations have also been studied, among which, overdispersed count-binary outcomes \citep{seyoum_joint_2017}, overdispersed count-ordinal outcome \citep{amini_longitudinal_2021}, and longitudinal and time-to-event data \citep{tsiatis_joint_2004, fitzmaurice_longitudinal_2008, efendi_joint_2013, njagi_flexible_2013}. The JSM, instead, has been limited to studying the association of longitudinal variables on time-to-event outcomes \citep{rizopoulos_joint_2012}. Here we present an adaptation of both models that can be used when the time-varying endogenous variable is continuous and the longitudinal outcome is of any kind (continuous, nominal, ordinal, bounded, etc.). For illustration, we will present simulations and applications motivated from DMD data, where the considered outcome is bounded. 

In practice, both JMM and JSM face significant computational challenges as the presence of multivariate longitudinal outcomes and random effects leads to log-likelihood functions involving high-dimensional integrals without closed-form solutions. Parameter estimation therefore requires numerical or simulation-based approximations, which can be computationally demanding or infeasible in complex settings \citep{pinheiro_approximations_1995}. Moreover, while existing R-packages for longitudinal modeling (\href{https://cran.r-project.org/web/packages/lme4/index.html}{lme4}, \href{https://cran.r-project.org/web/packages/nlme/index.html}{nlme} and \href{https://cran.r-project.org/web/packages/MCMCglmm/index.html}{MCMCglmm}) can be used to fit the JMM \citep{gomon_georgy_joint_2022}, the JSM cannot be estimated with even the most widely used joint modeling packages (\href{https://cran.r-project.org/web/packages/JM/index.html}{JM} and \href{https://cran.r-project.org/web/packages/JMbayes2/index.html}{JMbayes2}), as these are designed to fit JSM that include at least one time-to-event outcome, which is not our case.  
To address both the lack of software support for the JSM and the computational limitations of the parameters' estimation, we adopt the Bayesian estimation method called Integrated Nested Laplace Approximation \citep[INLA,][]{rue_approximate_2009, martins_bayesian_2013, gomez-rubio_bayesian_2021, niekerk_new_2021}. It is an approach that overcomes the issues coming from other Bayesian computation approaches such as Markov Chain Monte Carlo \citep[MCMC,][]{metropolis_equation_1953, hastings_monte_1970, geman_stochastic_1984}, which is used as an alternative to estimate GLMM (as in package \href{https://cran.r-project.org/web/packages/MCMCglmm/index.html}{MCMCglmm}). MCMC can be computationally expensive, especially in high-dimensional spaces, and it can lead to different results depending on the starting points. The INLA approach, instead, is deterministic, as it is based on optimization, and hence produces the same result each time it is run. Using INLA allows us to estimate joint models that can handle multiple types of outcomes, high-dimensional random effects, data with small sample size (typical for rare diseases) and short follow-up (typical for degenerative diseases), as the DMD data. 

One limitation is that neither the JMM nor the JSM provides a clear interpretation of the estimated associations between endogenous variables and the outcome. In JMM, conclusions about the association are based on the random effects' correlations. However, the interpretation in latent terms is not intuitive, especially when considering more than just random intercepts. Furthermore, this approach just gives information about the direction (positive or negative correlation) and magnitude of the association but does not provide an explicit quantification of the effect of the covariate's change in the outcome. JSM, on the other hand, facilitates a more direct description of the relationship between the outcome and the endogenous variable. However, when the outcome is not normally distributed, this relationship is only interpretable conditional on the random effects (subject-specific average). In such cases, JSM's fixed effects, similar to GLMMs, do not have a marginal interpretation (population average). Additionally, in both models, the association is interpreted in a cross-sectional way \citep{pepe_modeling_1997}, meaning it reflects the relationship between the two variables when measured at the same time. However, it may also be important to investigate the lagged effect, that is, how changes in one variable influence the other at later time points. 
We therefore propose an alternative method to quantify the association in both models, which helps to overcome all the issues just described and facilitate the communication of the results, particularly for researchers who are not experts in mathematics and statistics. That is, we provide a method for estimating a coefficient that has the same interpretation as a regression coefficient, thus indicating the mean change of the outcome, given a unit change in the covariate and all the other predictors fixed. This quantity also makes it possible to estimate not only the cross-sectional association (i.e., at the same time point) between the variables, but also the lagged effect of the endogenous variable on the outcome. To compute this coefficient, we employed Monte Carlo integration \citep{gelfand_sampling-based_1990}. A similar idea has been developed in the case of normal-binary outcomes \citep{delporte_joint_2022} or normal-ordinal outcomes \citep{delporte_joint_2025}. However, the method that we suggest is more general and can be applied across different types of outcomes, whereas their analytical solutions are specific for binary and ordinal outcomes. 

The paper is organized in the following manner. In Section \ref{DMD} we describe briefly the characteristics of the already mentioned motivating study, highlighting again the advantages of using joint models. In Section \ref{methods}, the methods are described. 
We present the mathematical form of the two joint models in Section \ref{sec2}, describing the process of derivation of the association quantities in Section \ref{ass}. In Section \ref{estimation} we describe the hierarchical structure and model formulation of JMM and JSM used by the INLA estimation method. Section \ref{simulation} presents the results of a simulation study conducted to evaluate the performance and accuracy of joint models when estimated using INLA. Finally, an application is shown in Section \ref{sec4}. To demonstrate the features of the proposed approach, we present analysis results from the DMD natural history study conducted at Leiden University Medical Center (LUMC).

\section{Motivating study}\label{DMD}
Duchenne Muscular Dystrophy \citep[DMD,][]{duan_duchenne_2021} is a genetic disorder characterized by progressive muscle wasting and weakness due to alteration of dystrophin, a protein essential for muscle integrity. Symptoms typically emerge between ages two and three, followed by progressive muscle degeneration. Hip and thigh muscles begin to weaken as early as five to seven years of age, and by early adolescence, most patients require full-time use of wheelchair. Respiratory and cardiac complications arise in the teens and early twenties, with most patients not surviving beyond their twenties. All of these stages in the advancement of the disease are summarized in functional scores, which are the results of physical tests performed by patients. 
However, this method of assessment is both time-consuming and motivation-sensitive. Thus, the goal of the study is to identify blood biomarkers, such as proteins, that are objective, easier to collect, and can more efficiently detect changes in disease progression.

\begin{figure}[t]
    \centering
    \begin{minipage}{0.49\textwidth}
        \centering
        \subfigure{
            \includegraphics[width=\linewidth]{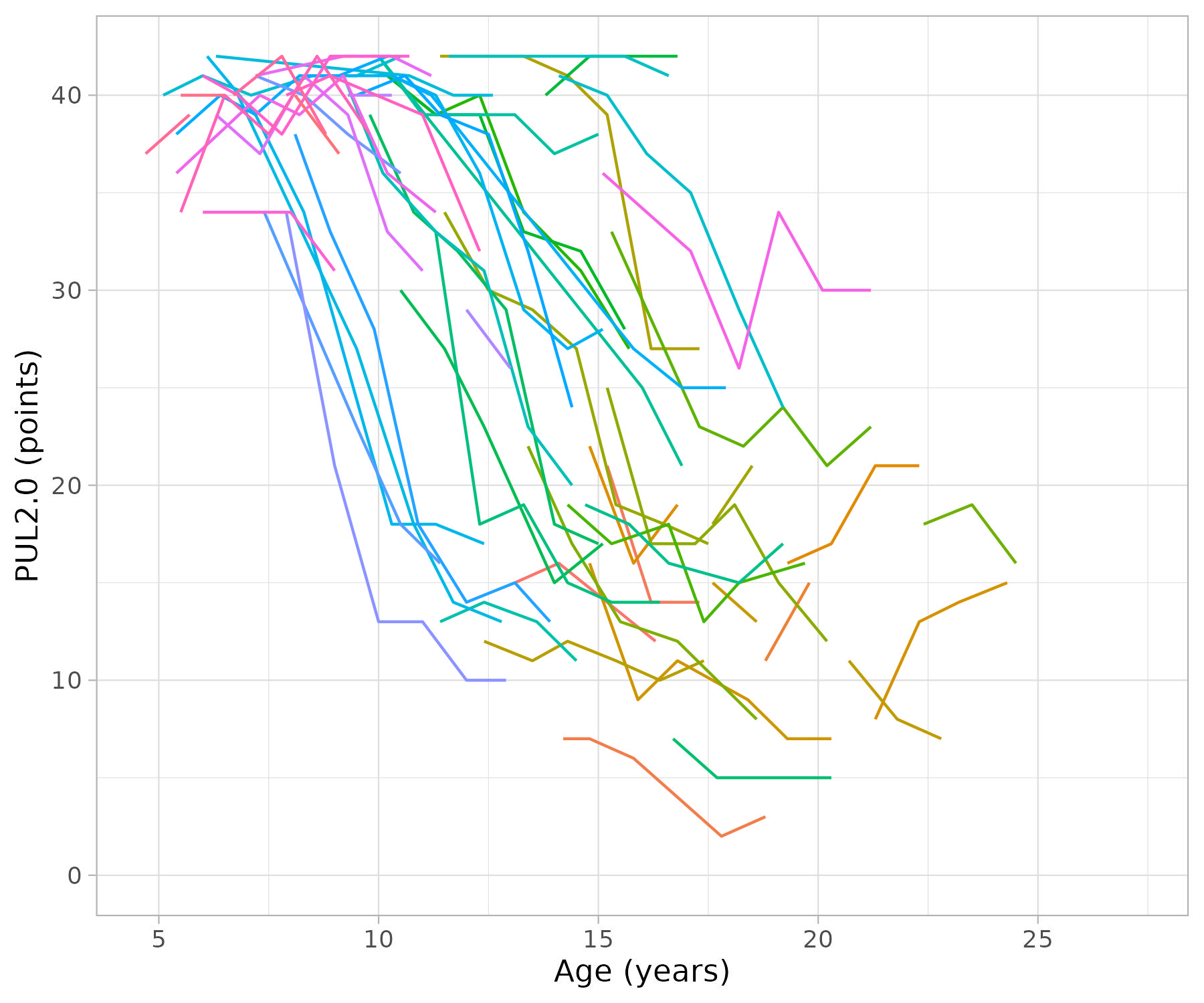}
        }
    \end{minipage}
    \hfill
    \begin{minipage}{0.49\textwidth}
        \centering
        \subfigure{
            \includegraphics[width=\linewidth]{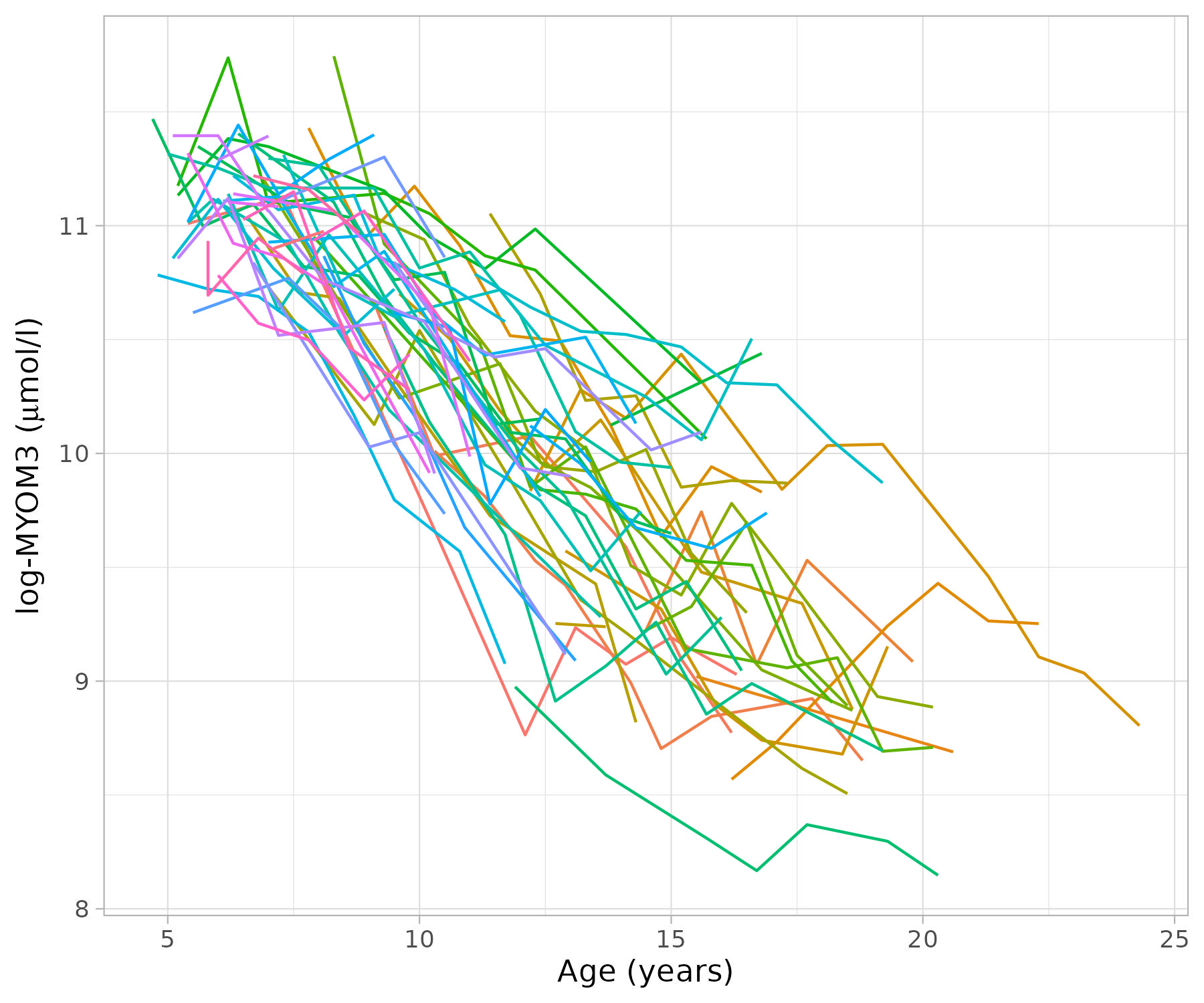}
        }
    \end{minipage}
    \caption{Longitudinal evaluation of PUL2.0 scores and MYOM3 protein levels across individual patients.}
    \label{spaghetti plots}
\end{figure}

The data available were collected from 65 DMD patients aged between 4.70 to 16.20 years at their first visit (mean: 8.054, SD: 3.076), who were followed up to 10 years (min: 2, max: 13, median: 8) at Leiden University Medical Center (LUMC). For our illustration, we selected one functional score: the Performance of Upper Limb 2.0 (PUL2.0) physical test. PUL2.0 is the total score of the upper limb performance test \citep{mayhew_performance_2020}. It collects information on several shoulder, elbow and hand movement tests, each of which is assigned a score. PUL2.0 is a bounded variable, with a range [0, 42] (Figure \ref{spaghetti plots}). Higher PUL2.0 scores reflect better upper limb function and greater independence, with scores around 42 indicating preserved function despite minor impairment. Lower scores, instead, indicate more severe impairment. As longitudinal endogenous variables, we selected five proteins already well known from previous publications: MYOM3, CNDP1, FABP1, LEP, MYL3, TTN. The goal was to study the association between these proteins and the PUL2.0 physical test. The time variable is age.

Considering this outcome, at some visits no blood sample was collected, so only the clinical test results were available, and vice-versa, sometimes no test was recorded in the dataset the day of the blood sample collection. Specifically, in $37\%$ of the considered samples the outcome was missing, but not the protein; and in $17\%$ of the samples the protein was missing and PUL2.0 was available. Therefore, using Joint Models we were able to save approximately $50\%$ of the available observations. Results are presented in Section \ref{sec4}.

\section{Methods}\label{methods}
The notation is presented in a general form, without reference to the motivating study, to facilitate the application of the analysis in other contexts.
We assume N independent subjects, indexed by $i = 1, \dots, N$, which are followed over time. Time can be continuous (e.g., age) or discrete (e.g., visit number); for generality, we assume a continuous time scale. For each subject, a time-varying outcome and a time-varying covariate are measured at potentially different sets of time points. Let $\mathbf{s}_i = (s_{i1}, \dots, s_{in_i})$ denote the observation times of the covariate for subject $i$, with $n_i$ repeated measurements, and $\mathbf{t}_i = (t_{i1}, \dots, t_{im_i})$ denote the observation times of the outcome, with $m_i$ repeated observations. The vectors $\mathbf{s}_i$ and $\mathbf{t}_i$ may differ in length and in timing, but coincide if the covariate and outcome are measured simultaneously for subject $i$. The observed covariate values of the $i$-th subject are $\mathbf{v}_i = (v_{i1}, \dots, v_{in_i})$, and the observed outcome values are $\mathbf{y}_i = (y_{i1}, \dots, y_{im_i})$. We indicate as $y_{ik}$, $k = 1, \dots, m_i$, the observed outcome value for the $i$th subject at time points $k$, and $v_{ij}$, $j = 1, \dots, n_i$, the observed covariate value for the same subject at time point $j$. Aggregating all the subjects, $\mathbf{v} = (\mathbf{v}_1, \dots, \mathbf{v}_N)$ and $\mathbf{y} = (\mathbf{y}_1, \dots, \mathbf{y}_N)$. \

The time-varying covariate $\mathbf{v}$ is defined as endogenous, meaning its current value at a specific time is not only associated with its previous values, but is further associated with previous values of the outcome. Mathematically speaking, $\mathbf{v}$ is endogenous if the exposure value at time point $j$, is not conditionally independent of the history of the response variable, given the history of the exposure process: $v_{ij} \not \perp \!\!\! \perp \{y_{i1}, y_{i2}, ..., y_{ih}; h < j\} \mid \{v_{i1}, v_{i2}, ..., v_{ih}; h < j\}$ \citep{qian_linear_2020}. If this dependence does not hold, the covariate is exogenous.\

In addition, we assume to observe $p$ time-fixed and/or exogenous covariates potentially associated with the outcome (e.g., treatment, sex, etc.) and other $q$ such covariates that could have an effect on the endogenous process (e.g., comorbidities, socioeconomic status, etc.). While it is possible that $p = q$ and the covariates overlap, for generality we treat them as distinct sets.

\subsection{Joint models of longitudinal outcome and endogenous time-varying covariates}\label{sec2}
To construct joint models, it is essential to define the processes of all involved variables. We assume only one continuous endogenous variable, modeled by a LMM; and one outcome that can be of any type, modeled by a GLMM. 
Let $\textbf{X}^{\text{v}}$ be the design matrix which specifies how time $\mathbf{s}$ and $q$ additional covariates influence the endogenous covariate process. Similarly, let $\textbf{X}^{\text{y}}$ be the fixed effects design matrix which specifies how time $\mathbf{t}$ and other $p$ covariates influence the outcome process. The matrices $\textbf{Z}^{\text{v}}$ and $\textbf{Z}^{\text{y}}$ are the corresponding random-effects design matrices. 
$\beta$ parameters denote fixed effects and $b$ parameters denote random effects. Superscripts distinguish between endogenous variable and outcome vectors of parameters, i.e., $\boldsymbol{\beta}^\text{v}$, $\boldsymbol{\beta}^\text{y}$, $\mathbf{b}^{\text{v}}$, $\mathbf{b}^{\text{y}}$. With these elements established, we can define the two models mentioned in the Introduction.

\subsubsection{Joint Mixed Model}\label{subsec1}
The mathematical form of the JMM assuming only one endogenous covariate and one outcome is:
\begin{equation}
\label{JMM}
\left\{\begin{aligned}
& \mathbf{v}_i = \mathbf{X}^{\text{v}}_i\boldsymbol{\beta}^\text{v} + \mathbf{Z}^\text{v}_i\mathbf{b}^\text{v}_i+\boldsymbol{\varepsilon}_i \\
& g(\boldsymbol{\mu}_i) = \boldsymbol{\eta}_i = \mathbf{X}^{\text{y}}_i\boldsymbol{\beta}^\text{y} + \mathbf{Z}^{\text{y}}_i\mathbf{b}^{\text{y}}_i
\end{aligned}\right.
\end{equation}
where $g$ is the link function, $\boldsymbol{\mu_i} = E[\mathbf{y}_i| \boldsymbol{\beta}^\text{y}, \mathbf{b}^{\text{y}}_i]$ is the mean of $\mathbf{y}_i$ conditional on the random effects and $\boldsymbol{\eta}_i$ is the linear predictor of $\mathbf{y}_i$. Here, we assume a normal distribution for the vector of error terms $\boldsymbol{\varepsilon}_i \sim \mathcal{N}(\boldsymbol{0}, \mathbf{\Sigma}_i = \sigma^{2}\textbf{I}_{n_i})$. This form of $\mathbf{\Sigma}_i$ implies the conditional independence assumption for the process of the endogenous covariate, which means that repeated measurements are conditionally independent given the random effects and that only the random effects capture the correlations within subject. \\
In this model, the association between the endogenous time-varying variable $\mathbf{v}$ and the outcomes $\mathbf{y}$ is measured via the random effects variance-covariance matrix $\mathbf{D}$, hence:
\begin{equation}
\label{eq:random effects}
\left[\begin{array}{l}
\mathbf{b}^{\text{v}}_{i} \\
\mathbf{b}^{\text{y}}_{i}
\end{array}\right] \sim \mathcal{N}\left(\mathbf{0},\left[\begin{array}{ll}
\mathbf{D}_{\text{v}} & \mathbf{D}_{\text{vy}} \\
\mathbf{D}_{\text{yv}} & \mathbf{D}_{\text{y}}
\end{array}\right]\right).
\end{equation}
If both processes have random intercepts and random slopes, $\mathbf{D}$ is a block matrix with submatrices the variance-covariance matrix of the random effects of the covariate $\mathbf{D}_{\text{v}}$, the variance-covariance matrix of the random effects of the outcome $\mathbf{D}_{\text{y}}$ and the covariance matrix $\mathbf{D}_{\text{vy}} =  \mathbf{D}_{\text{yv}}$ between them. If $\mathbf{D}_{\text{vy}}$ is null, the random effects are uncorrelated and there is no association between the response and the endogenous variables. \\
The entire association between the two variables is captured by the correlation of the random effects, that is, endogenous and response variables are independent conditional on random effects.

\subsubsection{Joint Scaled Model}\label{subsec2}
The JSM employs a scaling factor $\gamma$ to quantify the association between the endogenous time-varying variable $\mathbf{v}$ and the outcomes $\mathbf{y}$. The mathematical notation, assuming only one endogenous covariate and one outcome, is:
\begin{equation}
\label{JSM}
\left\{\begin{aligned}
& \mathbf{v}_i=\mathbf{m}_i +\boldsymbol{\varepsilon}_{i} \\
& \mathbf{m}_i = \mathbf{X}^{\text{v}}_{i} \boldsymbol{\beta}^{\text{v}}+\mathbf{Z}^{\text{v}}_{i} \mathbf{b}^{\text{v}}_{i} \\
& g(\boldsymbol{\mu}_i) = \boldsymbol{\eta}_i  = \mathbf{X}^{\text{y}}_i\boldsymbol{\beta}^\text{y} + \mathbf{Z}^{\text{y}}_i\mathbf{b}^{\text{y}}_i + \gamma \mathbf{m}_i
\end{aligned}\right.
\end{equation}
where $\mathbf{m}_i$ is the linear predictor of the endogenous time-varying variable for the $i$th subject. It is assumed that the random effects $\mathbf{b}^{\text{v}}_{i}$ and $\mathbf{b}^{\text{y}}_{i}$, and the error terms $\boldsymbol{\varepsilon}_{i}$, are independent and normally distributed, i.e., $\mathbf{b}^{\text{v}}_{i} \sim \mathcal{N}\left(\mathbf{0}, \mathbf{D}_\text{v}\right)$, $\boldsymbol{\varepsilon}_{i} \sim \mathcal{N}\left(\mathbf{0}, \mathbf{\Sigma}_i = \sigma^{2}\textbf{I}_{n_i}\right)$, $\mathbf{b}^{\text{y}}_{i} \sim \mathcal{N}\left(\mathbf{0}, \mathbf{D}_\text{y}\right)$.\\
The scaling factor $\gamma$ has the same interpretation as the fixed effects $\boldsymbol{\beta}^{\text{y}}$. In case of the identity link function, it expresses the mean change in the response variable for a unit change in the covariate model's linear predictor $\mathbf{m}_i$. Alternatively, it expresses how a unit change in the linear predictor $\mathbf{m}_i$ influences the transformed expectation of the outcome, given a specific subject. If $\gamma$ is zero, there is no association between the endogenous time-varying variable and the outcome.\\
The entire association between the two variables relies on the scaling factor, that is, random effects and errors are independent between the outcome and the endogenous variable. \\

\subsection{Association coefficient}\label{ass}
Interpreting the association in JMM (\ref{JMM}) is not trivial. The association relies on the correlations of the random effects, which can only provide information about the direction of the association (through their sign) as well as the order of magnitude of the association (high covariance indicates strong association). However, in some instances this information may not be sufficient. This is particularly relevant in medical and biological contexts. In these fields, it is crucial to quantify how much the expectation of one variable varies in reaction to expectation changes in another, in order to make informed decisions, such as selecting appropriate treatments or evaluating new therapies. It is also important to define and quantify potential lagged effects, as delayed responses between variables can carry significant implications for clinical interpretation. Moreover, outcomes may not always be normally distributed. In such cases, the association is no longer directly between the original values of the variables but rather between the original scale of the endogenous variable and the transformed mean (e.g., logit, probit, etc.) of the response variable (see for instance Section \ref{sec4}). As a result, interpretability becomes even less intuitive.

On the other hand, the interpretation of the association between the two variables in JSM (\ref{JSM}) is more straightforward. However, here as well, if the outcome variable is non-normal, the interpretation of the parameter $\gamma$ is expressed in terms of the effect (conditional on the random effects) of the endogenous variable directly on the transformed mean values of the outcome.

Nevertheless, a regression-type parameter interpretation is more appealing, both because it is easier to communicate and because it makes it possible to compare the results obtained with the two models. We define a point value $a$ of the endogenous variable $v_{ij}$, for which we want to quantify the effect on the expected value of the response variable $y_{ik}$. The resulting coefficient
\begin{equation}
\label{beta joint}
\beta_{joint}(a) = E\left[y_{ik} \mid v_{ij}  = a +1\right] - E\left[y_{ik} \mid v_{ij} = a\right]
\end{equation} 
expresses the change in expected outcome due to a unit increase in the endogenous covariate at value $a$. This coefficient idea aligns with the concept of Average Partial Effects \citep[APE, ][]{stoker_consistent_1986, klyne_average_2025}, where the focus is on defining how the expected outcome changes due to a small perturbation $\delta$ in a covariate within non-linear models. Here we set $\delta = 1$, as is most common, although in some clinical or biological contexts it may be more meaningful to consider alternative magnitudes of change. However, our proposal differs from APE in that the coefficient explicitly depends on the value of $a$. We consciously chose not to average out $a$, as doing so would obscure the fact that different levels of the endogenous variable can induce distinct effects on the outcome. This is particularly important when the conditional mean of $E[y_{ik} \mid v_{ij}]$ deviates from linearity, where averaging could mask meaningful heterogeneity in the association. \\
To estimate $\beta_{joint} (a)$, the marginal mean needs to be derived:
\begin{align}
E\left[y_{ik} \mid v_{ij} = a\right] 
&= \int_y y_{ik} f\left(y_{ik} \mid v_{ij}=a\right) \; d y_{ik} \notag \\
&= \int_y y_{ik} \; \frac{f\left(y_{ik}, v_{ij} = a\right)}{f(v_{ij} = a)} \; d y_{ik} \notag \\
&= \int_y y_{ik} \; \frac{f\left(y_{ik}, v_{ij} = a \mid \mathbf{b}_i \right) f(\mathbf{b}_i)}{f(v_{ij} = a)} \; d y_{ik} \notag \\
&= \int_b \int_y y_{ik} f\left(y_{ik} \mid \mathbf{b}_i\right) f\left(\mathbf{b}_i \mid v_{ij} = a\right) \; d y_{ik} \, d \mathbf{b}_i \label{alpha}
\end{align}
where $\mathbf{b}_i = (\mathbf{b}^{\text{v}}_i, \mathbf{b}^{\text{y}}_i)^{\top}$ represents the vector of random effects for $\mathbf{v}_i$ and $\mathbf{y}_i$, and $f(\mathbf{b}_i)$ denotes the joint multivariate distribution of the random effects. The integral is factorized in $\int_y y_{ik} f\left(y_{ik} \mid \mathbf{b}_i\right) d y_{ik}$, which is the mean of $y_{ik}$ conditional on the random effects $\mu_{ik} = g^{-1}(\eta_{ik})$ (\ref{JMM} and \ref{JSM}), and $f\left(\mathbf{b}_i \mid v_{ij}=a\right)$, which is the posterior distribution of the random effects. The estimation of this integral is described in Section \ref{ass_estimation}. If both variables are normal, the conditional mean of $y_{ik}$ given $v_{ij}$ can be analytically derived by a closed formula \citep{gomon_georgy_joint_2022, drikvandi_framework_2024}. The steps in deriving (\ref{alpha}) are given in detail in Section 1 and 2 of the \href{Supplementary_material/main.tex}{Supplementary Material}, including the special case where both variables involved are normally distributed.\

The coefficient $\beta_{joint}$ (\ref{beta joint}) depends on the time of collection of both the endogenous variable $\mathbf{s}$ and the response variable $\mathbf{t}$. When we assume the same time $s_{ij} = t_{ik}$, we capture the cross-sectional effect of the endogenous variable on the outcome, thus the change in mean outcome for a unitary increase in the endogenous covariate given the same time point.
This feature of $\beta_{joint}$ is particularly relevant to studies on degenerative disease, such as DMD (Section \ref{DMD}), where monitoring changes over time is really crucial. Understanding how the association between the biomarker and the physical test develops with age may help to make better decisions. For instance, there may be no association between the two variables at early ages, making it not feasible to draw conclusions based on the patient's biomarker levels. Nevertheless, as the disease progresses and becomes more severe, the association may strengthen with age, allowing for new conclusions to be drawn. This concept will be clarified in Section \ref{sec4}.  
It is also possible to assume different time points for the two variables. In this scenario, $\beta_{joint}$  measures the lagged effect.
If the time it takes for one variable to influence another is not taken into account, studies may draw the wrong conclusions. By assuming two different time points, it is possible to measure how much past values of the covariate associate with the future outcome. 
Alternatively, we can obtain the average association over time, resulting in a time-invariant association coefficient. \

\section{Estimation and inference}\label{estimation}
\subsubsection{Estimation of the models}\label{INLA_estimation}
In univariate generalized linear mixed models, parameter estimation is achieved through maximum likelihood estimation on approximations of the marginal likelihood. The same approach can theoretically be applied to the joint models in (\ref{JMM}) and (\ref{JSM}) to estimate parameters from the joint marginal distribution:
\begin{equation}
\label{int joint}
    f(\mathbf{v}_i, \mathbf{y}_i) = \int f(\mathbf{v}_i, \mathbf{y}_i \mid \mathbf{b}_i)f(\mathbf{b}_i) \; d \mathbf{b}_i \\
\end{equation}
with $\mathbf{b}_i = (\mathbf{b}^{\text{v}}_i, \mathbf{b}^{\text{y}}_i)^{\top}$ being the joint vector of random effects. However, when one of the outcomes is non-normally distributed, the integral (\ref{int joint}) does not have a closed-form solution and must be approximated using numerical methods. Additionally, the computational burden required to estimate the integral increases exponentially with the number of random effects in the model. 

To address these challenges, we decided to proceed with a Bayesian estimation approach, namely the Integrated Nested Laplace Approximation (INLA). INLA offers a fast and deterministic alternative to the commonly used Markov Chain Monte Carlo (MCMC) methods. INLA models are three-stage hierarchical models, and each stage must be specified for the analysis. In the first stage, we define the distribution of the observed data, e.g., the likelihood; in the second stage, the prior distributions for the parameters (fixed and random effects); and in the third stage, the prior distribution for the hyperparameters, including the precision matrix of the random effects, the precision of the error terms, and the scaled parameter in the JSM. One of the most important aspects for correctly fitting joint models with INLA is selecting the appropriate priors, particularly for the hyperparameters. By default, INLA uses non-informative hyperpriors, but in scenarios with a very low number of observations, such as the motivating study detailed in Section \ref{DMD}, these may not be ideal. Therefore, we adjusted some of the initial hyperprior parameters to make them more informative. We based our decisions on the findings from our simulation studies; results are shown in Section 3.1 of the \href{Supplementary_material/main.tex}{Supplementary Material}. \

For the precision matrix of the random effects, we assumed the prior $\mathbf{D}^{-1} \sim Wishart_p(r, \mathbf{R}^{-1})$, where $p$ is the number of random effects in the model, $r$ is the number of degrees of freedom which depends on $p$ e.g., if $p = 2$ then $r = 4$, if $p = 4$ then $r = 11$ \citep{gomez-rubio_bayesian_2021}, and $\mathbf{R} = \text{diag}(\hat{\sigma}_1, \dots, \hat{\sigma}_p)$ is the scaled matrix. The default option assumes an identity matrix, but we propose an informative alternative where the diagonal of $\mathbf{R}$ contains the maximum-likelihood estimates of the standard deviations of the random effects $\hat{\sigma}_1, \dots, \hat{\sigma}_p$. 
Other hyperparameters of the model are the scaled parameter $\gamma$ in the JSM (\ref{JSM}) and the precision of the error terms $\boldsymbol{\varepsilon}_i$ on the endogenous process. For the former we assume $\gamma \sim \mathcal{N}(0, 1/0.0001)$, for the latter $1/\sigma^{2} \sim Ga(1, 0.00005)$.
For the fixed effects of the two processes, by default, the prior is non-informative with a wide variance $\beta \sim \mathcal{N}\left(0, 1/0.001\right)$. The random effects and the error terms both follow a Gaussian distribution: $\mathbf{b}_i \sim \mathcal{N}\left(\mathbf{0}, \mathbf{D}^{-1}\right)$ and $\boldsymbol{\varepsilon}_i \sim \mathcal{N}\left(0, \log(1/\sigma^2)\right)$.

INLA provides full Bayesian inference by returning marginal posterior distributions for all model parameters, including fixed effects and random effects. However, hyperparameters, such as variances and covariances, are not returned directly. Instead, INLA uses internal parameterizations: variances are represented through their inverses, known as precisions, and covariances are replaced by correlations. All posterior distributions are summarized through key statistics such as means, standard deviations, and credible intervals. Additionally, model fit and complexity can be assessed using Bayesian evaluation methods. The INLA output includes several such measures. The marginal likelihood \citep{rue_approximate_2009}, which quantifies how likely the observed data are under a given model; or the Bayesian information-based criteria  Deviance Information Criterion \citep[DIC,][]{gelman_bayesian_1995, rue_approximate_2009} and Watanabe-Akaike Information Criterion \citep[WAIC,][]{gelman_bayesian_1995}, which evaluate the model goodness of fit while adjusting for the order of complexity. 
DIC and WAIC can be computed either pointwise, providing values for each individual observation, or aggregated to obtain overall measures of model fit. Since both the JMM and JSM are models for both the outcome and the endogenous variable, this decomposition allows us to separately assess the overall model performance and the model’s ability to fit the outcome component conditional on the observed endogenous variable and the estimated parameters.  

\subsubsection{Estimation of the association coefficient}\label{ass_estimation}
In this section, we describe the procedure used to estimate the association coefficient, highlighting the computational challenges and the adopted numerical strategy. The procedure for estimating the integrals in both the JMM and JSM cases is identical. The integral in (\ref{alpha}) cannot be derived analytically because it depends on the factorization of two distributions that are neither of the same type nor conjugate: the conditional mean of the outcome given the random effects and the conditional distribution of the random effects given the endogenous variable. The  conditional mean of the outcome given the random effects can be specified as: 
\begin{equation} \label{cond mean}
   \int_y \mathbf{y}_i f\left(\mathbf{y}_i \mid \mathbf{b}_i\right) d \mathbf{y}_i = E[\mathbf{y}_i \mid \mathbf{b}_i] =g^{-1}(\boldsymbol{\eta}_i), 
\end{equation}
while the analytical specification of the conditional distribution of the random effects given the endogenous variable is:
$$
f(\mathbf{b}_i, \mathbf{v}_i) \sim \mathcal{N} \left(\boldsymbol{\mu} = \left[\begin{array}{l}
\boldsymbol{\mu}_\text{b}\\
\boldsymbol{\mu}_\text{v}
\end{array}\right], \mathbf{\Omega} = \left[\begin{array}{cc}
\mathbf{D} & \mathbf{\Omega}_{\text{bv}}\\
\mathbf{\Omega}_{\text{bv}} &  \text{var}(\mathbf{v}_i)
\end{array}\right]
\right) $$
\begin{equation} \label{cond b}
f(\mathbf{b}_i\mid v_{ij} = a) \sim \mathcal{N} \left(\boldsymbol{\mu}_\text{b} + \mathbf{\Omega}_{\text{bv}}  \text{var}(\mathbf{v}_{i})^{-1}(a - \boldsymbol{\mu}_\text{v}), \mathbf{D} - \mathbf{\Omega}_{\text{bv}}  \text{var}(\mathbf{v}_{i})^{-1} \mathbf{\Omega}_{\text{bv}}\right).
\end{equation}
However, the models (\ref{JMM}) and (\ref{JSM}) were estimated using the Bayesian estimation method INLA, which provides univariate marginal posterior distributions for all parameters and hyperparameters involved, along with summary statistics such as mean, mode, and quantiles. Since the integral is defined solely by conditional distributions, the INLA output cannot be easily used to address the estimation problem. As a result, the Bayesian results were combined with a frequentist approach, employing Monte Carlo integration to numerically approximate the integral. All involved distributions (\ref{cond mean}) and (\ref{cond b}) can be specified analytically. The mode of the univariate marginal posterior distributions was used as point estimates for the parameters involved, denoted as $\hat{\boldsymbol{\theta}} = (\hat{\boldsymbol{\beta}^{\text{v}}}, \hat{\boldsymbol{\beta}^{\text{y}}}, \hat{\mathbf{b}}_i, \hat{\boldsymbol{\varepsilon}_i}, \hat{\textbf{D}}, \hat{\sigma})$. With this information, the random effects were integrated out from the integrand, allowing for a numerical approximation of the association coefficient. For more details on the integral formulation and how the Monte Carlo approximation method was employed, see Section 1 and 2 in the \href{Supplementary_material/main.tex}{Supplementary Material}. Following this approach, $E\left[y_{ik} \mid v_{ij} = a\right]$ and $E\left[y_{ik} \mid v_{ij} = a + 1\right]$ were estimated, and the association coefficient $\beta_{joint}$ (\ref{beta joint}) was derived.\

An estimate of the uncertainty of $E\left[y_{ik} \mid v_{ij} = a\right]$ was also derived. This expectation is a function of the parameters and hyperparameters $\hat{\boldsymbol{\theta}} = (\hat{\boldsymbol{\beta}^{\text{v}}}, \hat{\boldsymbol{\beta}^{\text{y}}}, \hat{\mathbf{b}}_i, \hat{\boldsymbol{\varepsilon}_i}, \hat{\textbf{D}}, \hat{\sigma})$, which are considered random variables because their values depend on the specific random samples observed. Consequently, $E\left[y_{ik} \mid v_{ij} = a\right]$ is also a random variable, with its randomness coming from the parameters and hyperparameters. Thus, to estimate the uncertainty, we need to account for the uncertainty of $\hat{\boldsymbol{\theta}}$. The variability of the parameters $(\hat{\boldsymbol{\beta}^{\text{v}}}, \hat{\boldsymbol{\beta}^{\text{y}}}, \hat{\mathbf{b}}_i, \hat{\boldsymbol{\varepsilon}_i})$ in $\hat{\boldsymbol{\theta}}$ arises from the variability of the hyperparameters $\hat{\textbf{D}}, \hat{\sigma}$. Therefore, to estimate the uncertainty of $E\left[y_{ik} \mid v_{ij} = a\right]$, we just need to consider uncertainty of the hyperparameters $\hat{\textbf{D}}, \hat{\sigma}$. To achieve this, we augment the Monte Carlo integration approximation of integral (\ref{alpha}), by adding a new layer of resampling. We sample from the approximate joint posterior distribution of the hyperparameters $\hat{\textbf{D}}, \hat{\sigma}$ using a function available in INLA (\textit{inla.hyperpar.sample}). To see how the integration estimation is done on top on the Monte Carlo approximation refer to Figure 2 in the \href{Supplementary_material/main.tex}{Supplementary Material}; it is a schematic representation of the estimation process of both $\hat{\beta}_{joint}$ (\ref{beta joint}) and its variance $var(\hat{\beta}_{joint})$.

\section{Simulation} \label{simulation}
We conducted a simulation study to evaluate the performance of joint models and assess the accuracy of the results when using INLA as method of estimation. Specifically, we examined how the performance of the two joint models varies with the amount of observations available. Additional simulation results can be found in the Section 3 of the \href{Supplementary_material/main.tex}{Supplementary Material}. 

\subsection{Data generation mechanism}
The simulation study design was inspired by the motivating study described in Section \ref{DMD}. We considered two scenarios: one with only $N=65$ individuals and another with $N=200$. The generating mechanism for both scenarios is the same. \\
We simulated a maximum of 12 repeated measurements per individual, which should emulate age, ranging from 3.0 to 27.0. The only predictor included in the simulated models was time. The data were unbalanced, with a rate of missing values in the outcome of $28\%$ and $45\%$ missing values in the covariate. Data were simulated according to either a JMM or a JSM. For each joint model, we generated $B=1000$ samples, resulting in a total of $4B$ simulated datasets: $2B$ in the scenario with $N=65$ individuals, and the other half in the scenario with $N=200$ individuals. We considered one bounded outcome $\mathbf{y}$ with values in the range $(0,1)$ and one continuous endogenous variable $\mathbf{v}$.
The parameters utilized to simulate from the two joint models were:
$$
\left[\begin{array}{l}
\boldsymbol{\beta}^{\text{v}}_{i} \\
\boldsymbol{\beta}^{\text{y}}_{i}
\end{array}\right] = \left[\begin{array}{c}
12.108\\ -0.166\\ 4.666\\ -0.278
\end{array}\right],
\quad
\left[\begin{array}{l}
\mathbf{b}^{\text{v}}_{i} \\
\mathbf{b}^{\text{y}}_{i}
\end{array}\right] \sim \mathcal{N}\left(\mathbf{0},\left[\begin{array}{cccc}
0.243 & -0.019 & 0.654 & -0.056\\
  & 0.004 & -0.032 & 0.005\\
 &  & 6.004 & -0.38 \\
 &  & & 0.042\\
\end{array}\right]\right), 
\quad
\boldsymbol{\varepsilon}_{i} \sim \mathcal{N}\left(\mathbf{0}, 0.22\textbf{I}_{n_i}\right).
$$ In the JMM, all random effects are jointly distributed, whereas in the JSM, the random effects of the two variables are assumed to be independent. As a result, the form of the variance-covariance matrix of the random effects differs between the two models. However, the parameters of variances and covariances remain the same in both simulations.\\ 
The endogenous variable was specified as a linear combination of the fixed and random effects parameters with time, following the formulations in (\ref{JMM}) and (\ref{JSM}). The outcome was randomly generated from a beta distribution $\mathbf{y}_i \sim Beta(\boldsymbol{\mu_i}, \phi = 32.77)$, where $\boldsymbol{\mu}_i = g^{-1}(\boldsymbol{\eta}_i) = expit(\boldsymbol{\eta}_i)$ and $\phi$ is the precision value of the beta distribution.\\
The scale parameter $\gamma$ of the JSM was assumed to be equal to 2.57. \\
Both models have been estimated assuming the priors and hyperpriors distributions described in Section \ref{INLA_estimation}.

\subsection{Results}
\begin{figure}[t]
    \centering
    \begin{minipage}{0.49\textwidth}
        \centering
        \subfigure[Joint Mixed Model]{
            \includegraphics[width=\linewidth]{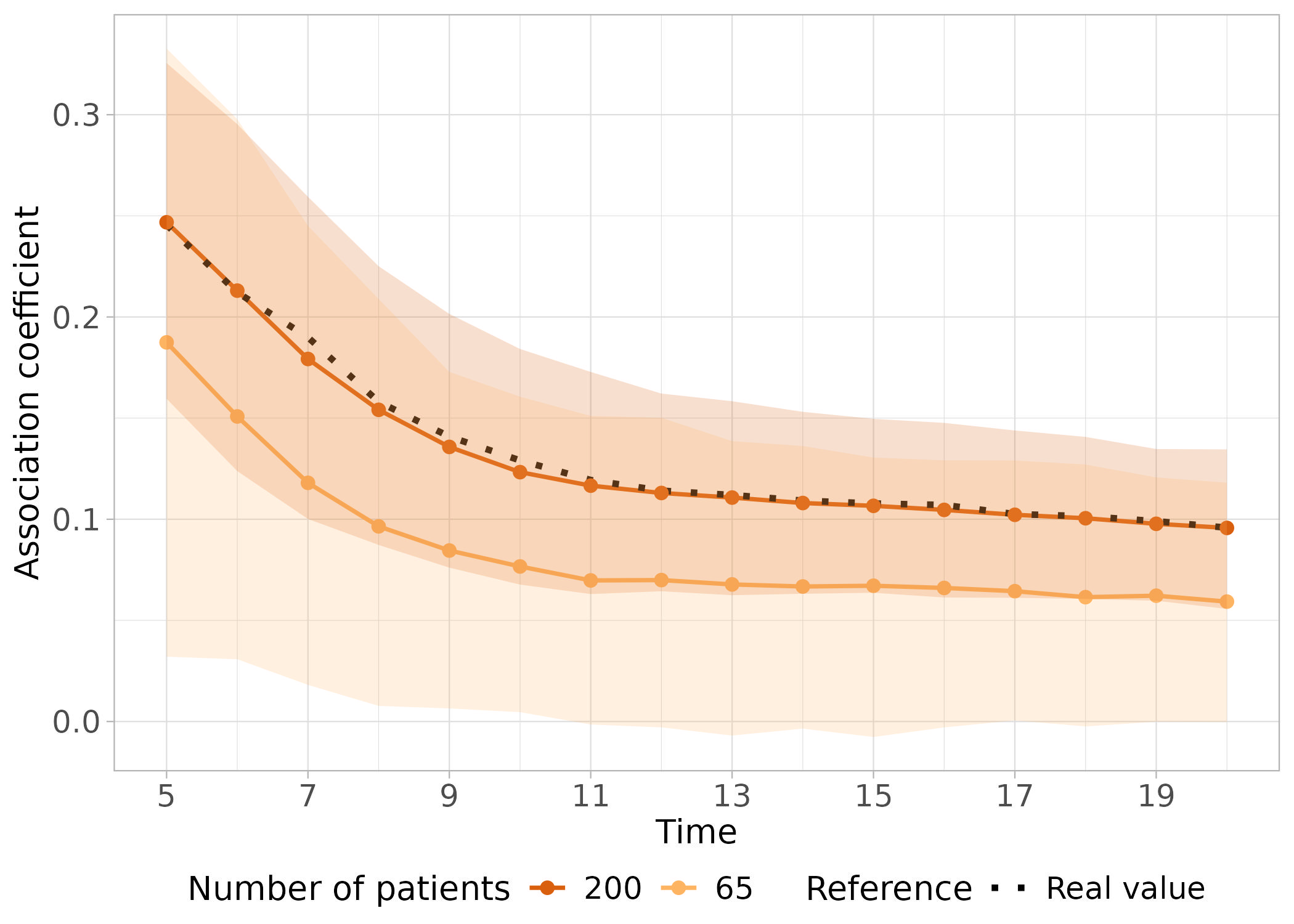}
        }
    \end{minipage}
    \hfill
    \begin{minipage}{0.49\textwidth}
        \centering
        \subfigure[Joint Scaled Model]{
            \includegraphics[width=\linewidth]{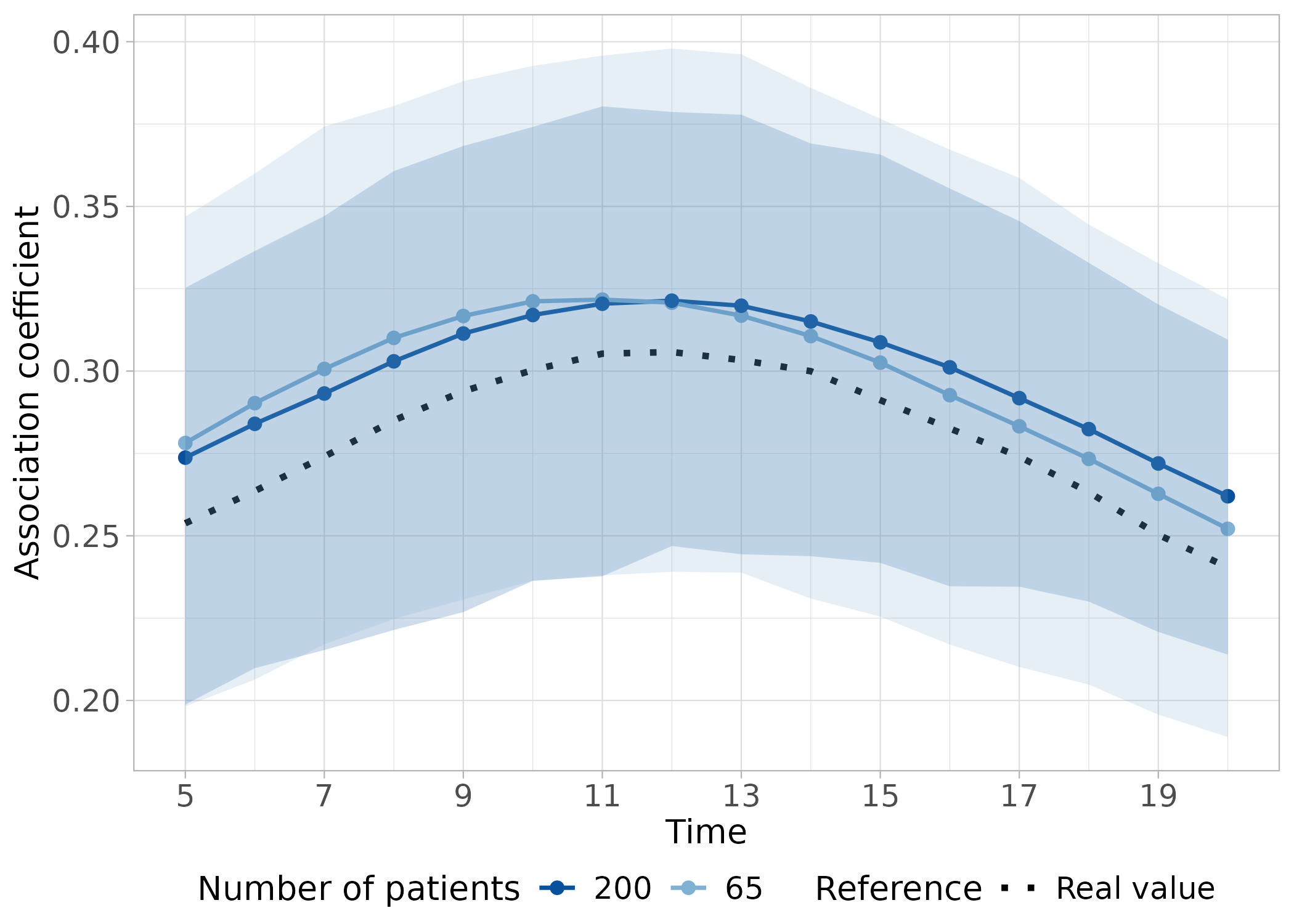}
        }
    \end{minipage}
    \caption{Evaluation of the estimation of the cross-sectional association between the endogenous covariate and the outcome, assuming the base value $a$ set equal to 9. Here the time variable emulates age. The solid lines represent the mean estimated association coefficients across $B = 1000$ simulations of the JMM (orange) and JSM (blue), for sample sizes $N = 65$ and $N = 200$. These are compared against the true value, indicated by the dotted black line. The figure also displays the corresponding 90\% credible intervals.}
    \label{figsim}
\end{figure}

\begin{figure*}[!t]%
\centering
\includegraphics[width=0.70\linewidth]{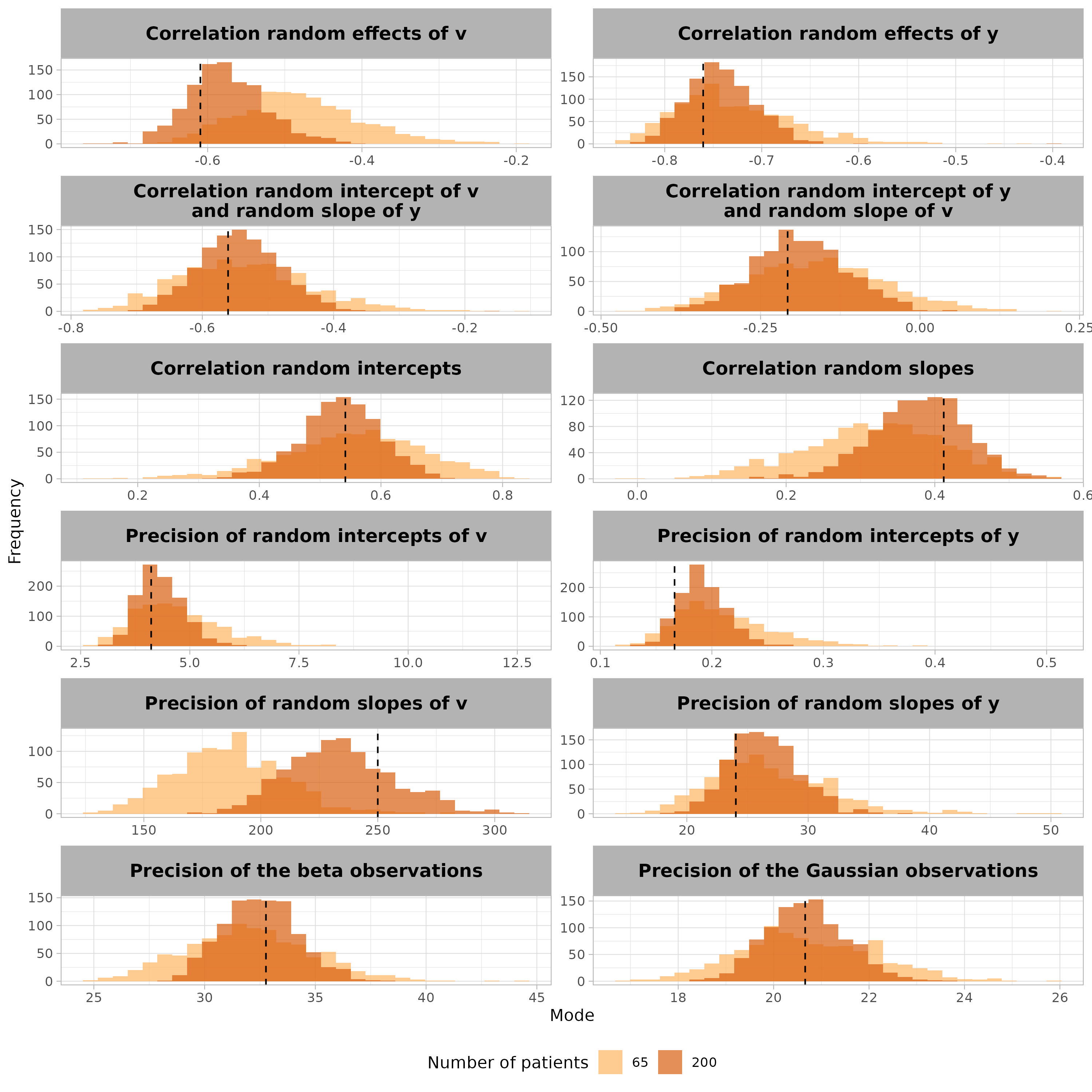}
\caption{Comparison of JMM estimates for sample sizes $N = 65$ and $N = 200$, stratified by hyperparameter. The histograms display the posterior modes of the hyperparameters from $B = 1000$ simulated datasets. The figure includes the variances and covariances of the random effects, the precision parameter $\phi$ of the Beta distribution governing the outcome, and the precision of the error terms $\epsilon_i$ associated with the endogenous variable.}
\label{fig_sim_jmm}
\end{figure*}

Figure \ref{figsim} presents the average association coefficients (\ref{beta joint}) estimated from $B = 1000$ replications of the JMM and JSM, applied to datasets with $N = 65$ and $N = 200$ individuals, along with the corresponding 90\% credible intervals. For illustration purposes, we show the results of (\ref{beta joint}) assuming the base value $a$ equal to the sample mean value 9. For smaller sample sizes, the JMM tends to underestimate the association coefficients, and although the credible intervals include the true values, their substantial width indicates poor estimations' precision. As the sample size increases to 200, both bias and variability decrease, and the estimates approach the true association values. 
Further insight is provided in Figure \ref{fig_sim_jmm}, which displays histograms of the posterior modes for each hyperparameter. When $N = 65$, the precision of the random slopes of $\mathbf{v}_i$ is notably underestimated. Consequently also the correlation between the random effects of $\mathbf{v}_i$ is overestimated, and the correlation between the random slopes of $\mathbf{v}_i$ and $\mathbf{y}_i$ is underestimated. Finally, the precision of the random intercept of $\mathbf{y}_i$ is overestimated. These estimation issues persist despite the use of informative hyperpriors for the variance-covariance matrix, typically employed to compensate for limited data, indicating that in this case, the added prior information is still insufficient to offset the lack of observational data and the model’s complexity. A sensitivity analysis comparing informative and non-informative priors is provided in the \href{Supplementary_material/main.tex}{Supplementary Material}. Differently, when $N = 200$, the hyperparameter estimates are much closer to the true values (dotted lines). Because the association coefficients (\ref{beta joint}) depend on these hyperparameters, their improved estimation directly enhances the accuracy of the JMM. 

\begin{figure*}[!t]%
\centering
\includegraphics[width=0.70\linewidth]{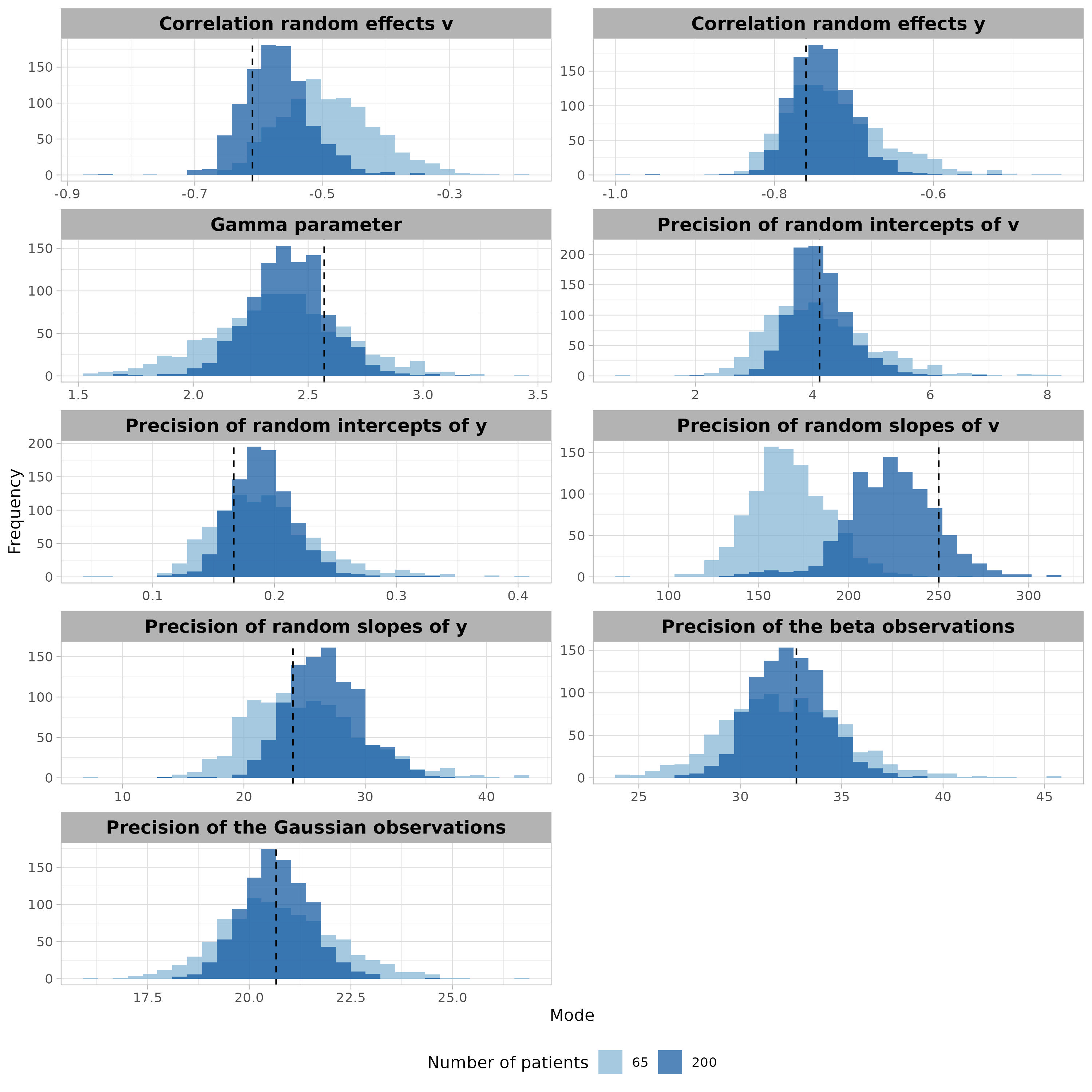}
\caption{Comparison of JSM estimates for sample sizes $N = 65$ and $N = 200$, stratified by hyperparameter. The histograms display the posterior modes of the hyperparameters from $B = 1000$ simulated datasets. The figure includes the variances and covariances of the random effects, the precision parameter $\phi$ of the Beta distribution governing the outcome, the precision of the error terms $\epsilon_i$ associated with the endogenous variable, and the scaling parameter $\gamma$.}
\label{fig_sim_jsm}
\end{figure*}

In contrast, fitting the JSM with either $N = 65$ or $N = 200$ individuals does not appear to substantially affect the results (Figure \ref{figsim}). In both scenarios, the estimated association coefficient remains close to the true value, but slightly overestimated, with only a slight increase in variability observed when $N = 65$. Overall, the JSM demonstrates greater stability compared to the JMM. However, Figure \ref{fig_sim_jsm} suggests that the estimation of certain hyperparameters still benefits from a larger sample size. Notably, the precision of the random slopes for $\mathbf{v}_i$ is substantially underestimated when $N=65$, and the correlation between the random effects of $\mathbf{v}_i$ tends to be overestimated. The parameter gamma also seems slightly underestimated.

The relatively lower sensitivity of the association coefficients (\ref{beta joint}) to sample size in the JSM can be attributed to the model’s simpler structure, which involves fewer hyperparameters due to the absence of correlations between random effects. In contrast, the JMM requires the estimation of numerous hyperparameters, which necessitates a larger number of observations to achieve correct estimates.

\section{Analysis of the data}\label{sec4}
In this section, the DMD data described in Section \ref{DMD} are analysed. We begin by assessing the association between the functional score PUL2.0 and log-MYOM3, followed by a presentation of results for the remaining proteins at the end of the section. The analysis was conducted using both a JMM and a JSM. 
For the protein component of the joint models, the same linear mixed model was defined in both JMM and JSM:
$$
\log(v_{ij}) = m_{ij} + \varepsilon_{ij}= \beta_0^\text{v} + b_{0i}^\text{v} + (\beta_1^\text{v} + b_{1i}^\text{v})s_{ij}+\varepsilon_{ij},
$$
where $s_{ij}$ is the age of patient $i$ at the visit $j$ of collection of blood measurements.
For the outcome component, instead, we assumed a GLMM with beta distribution and with logit link function:
\begin{align*}
\log\left(\frac{\boldsymbol{\mu}_{ik}}{1 - \boldsymbol{\mu}_{ik}}\right) 
  &= \boldsymbol{\eta}_i 
   = \beta_0^\text{y} + b_{0i}^\text{y} + (\beta_1^\text{y} + b_{1i}^\text{y}) t_{ik}, 
   && \text{(JMM)} \\
\log\left(\frac{\boldsymbol{\mu}_{ik}}{1 - \boldsymbol{\mu}_{ik}}\right) 
  &= \boldsymbol{\eta}_i 
   = \beta_0^\text{y} + b_{0i}^\text{y} + (\beta_1^\text{y} + b_{1i}^\text{y}) t_{ik} + \gamma m_{ik}, 
   && \text{(JSM)}
\end{align*}
where $t_{ik}$ represents the age of patient $i$ at the $k$-th physical test appointment, and $m_{ik}$ corresponds to the linear predictor of the protein at time $t_{ik}$.
The Beta distribution is defined in a unit interval (0, 1) and is often used in the presence of a continuous but bounded dependent variable \citep{bonat_likelihood_2014, figueroa-zuniga_mixed_2013}. Since the PUL2.0 outcome is defined between [0, 42], we implemented the model on the transformed variable $y_{ik}^* = y_{ik}/42$, with $y_{ik}$ being the single PUL2.0 observation. A test value equal to 1 indicates that the patient can complete 100\% of all tasks, while a result equal to 0 means that the patient is no longer ambulatory and constantly needs external support. Here, $\mu_{ik}$ is the expected transformed values of the PUL2.0 test $y_{ik}^*$. The Beta distribution, combined with the logit transformation, provides a suitable framework for modeling the S-shaped nature of the PUL2.0 outcome (Figure \ref{spaghetti plots}) and helps to mitigate the ceiling effect. \\
For both variables, only age was considered as predictor, and we assumed both random intercepts and random slopes. In the JMM, the random effects were assumed to follow a multivariate normal distribution with an unstructured variance–covariance matrix:
$$
\left[\begin{array}{l}
\mathbf{b}^{\text{v}}_{i} \\
\mathbf{b}^{\text{y}}_{i}
\end{array}\right] \sim \mathcal{N}\left(\mathbf{0},\left[\begin{array}{cccc}
\sigma^2_{1} & \rho_{12}\sigma_1\sigma_2 & \rho_{13}\sigma_1\sigma_3 & \rho_{14}\sigma_1\sigma_4\\
  & \sigma^2_{2} & \rho_{23}\sigma_2\sigma_3 & \rho_{24}\sigma_2\sigma_4\\
 &  & \sigma^2_{3} & \rho_{34}\sigma_3\sigma_4\\
 &  & & \sigma^2_{4}\\
\end{array}\right]\right),
$$ where $\sigma$ are the standard deviations of the random effects, and $\rho$ are the corresponding correlations. In the JSM, the same specification is adopted except that the random effects of $\mathbf{y}_i$ and $\mathbf{v}_i$ were assumed independent. This corresponds to constraining $\rho_{13}, \rho_{14}, \rho_{23}, \rho_{24}$ to zero, i.e., imposing a block-diagonal structure on the variance-covariance matrix.

In Table \ref{tab1} are shown the estimates of the two models. As the central metric of the parameters, we considered the mode, and the uncertainty of the parameters was assessed through credible intervals. Estimates for JMM and JSM are quite similar. In both cases, the standard deviations of the random effects indicate variability among patients for either the protein (JMM: $\sigma_1 = 0.452$; JSM: $\sigma_1 = 0.666$) and the outcome measures (JMM: $\sigma_3 = 2.418$; JSM: $\sigma_3 = 2.367$), primarily captured by the random intercepts. Notably, for the transformed-PUL2.0, a strong negative correlation exists between the random effects (JMM: $\rho_{34} = -0.814$; JSM: $\rho_{34} = -0.773$), meaning that patients with higher intercept tend to have flatter slopes (or vice versa). In other words, patients with higher PUL2.0 test scores tend to decline more slowly, as they manifest better functional performance.

Looking more specifically at the JMM results, there appears to be a strong positive correlation between the random intercepts of the two variables ($\rho_{13} = 0.648$). The random intercepts reflect how each patient's average level of a variable deviates from the population mean. A high correlation between them suggests that patients with higher average PUL2.0 scores also tend to have higher average protein values. Thus we expect that the higher the protein value, the better the performance of the physical test. There is also moderate positive correlation between the random slopes ($\rho_{24} = 0.423$), indicating wheatear or not patients that declines more rapidly in the transformed-PUL2.0 tend to decline rapidly in protein levels. There is as well a moderate correlation between the random intercepts of the protein and random slopes of the outcome ($\rho_{14} = -0.64$). This suggests that patients with higher average MYOM3 protein levels tend to experience a slower decline in transformed-PUL2.0 progression.

Beyond interpreting random-effects correlations, the JSM and its scaled parameter provide further insights into the association between the protein and the PUL2.0 outcome. Conditional on the other covariates and random effects, the interpretation of the parameter $exp(\gamma)$ can be expressed in terms of the effect of the protein on the odds ratio $\frac{\mu_{ik}}{1 - \mu_{ik}}$, $\mu_{ik} = E[y_{ik}|\mathbf{b}_i^\text{y}]$. The $\gamma = 2.55$ parameter estimate (Table \ref{tab1}) suggests that a unit increase in the linear predictor of the protein corresponds to better performance in the test. More precisely, for a given subject, increasing log-MYOM3 by one unit results in a 2.55-unit increase in the log-odds of the outcome.
However, interpreting the association in terms of log-odds remains complex and not really practical. Additionally, the fixed effects estimates (as $\gamma$) in GLMM quantify the effect of the variable conditionally on the random effects rather than marginally \citep{molenberghs_models_2005}, which is not always desirable.

\begin{table*}[t]
\caption{The hyperparameters estimates of the joint models.\label{tab1}}
\tabcolsep = 10pt
\renewcommand{\arraystretch}{1.2}
\begin{tabular*}{\textwidth}{@{\extracolsep{\fill}}lccccc@{\extracolsep{\fill}}}
\toprule%
& \multicolumn{2}{c}{JMM} & \multicolumn{2}{c}{JSM} \\
\cline{2-3}\cline{4-5}%
Parameters & Mode$^{1}$ & CI$^{2}$ & Mode$^{1}$ & CI$^{2}$ \\
\midrule
$\sigma_1$ & 0.452 & (0.295, 0.789) & 0.666 & (0.502, 0.931) \\
$\sigma_2$ & 0.067 & (0.052, 0.088) & 0.138 & (0.117, 0.168)\\
$\sigma_3$ & 2.418 & (1.803, 3.371) & 2.367 & (1.814, 3.291) \\
$\sigma_4$ & 0.200 & (0.150, 0.279) & 0.239 & (0.194, 0.308) \\[5pt]

$\rho_{12}$ & -0.611 & (-0.789, -0.245) & -0.408 & (-0.592, -0.147)\\
$\rho_{13}$ & 0.648 & (0.056, 0.853) & - & - \\
$\rho_{14}$ & -0.64 & (-0.840, -0.130) & - & -\\
$\rho_{23}$ & -0.218 & (-0.594, 0.258) & - & -\\
$\rho_{24}$ & 0.423 & (-0.041, 0.702) & - & -\\
$\rho_{34}$ & -0.814 & (-0.891, -0.630) & -0.773 & (-0.859, -0.612)\\[5pt]

$\gamma$ & - & - & 2.55 & 1.70 - 3.40\\
$exp(\gamma)$ & - & - & 14.01 & 5.48 - 29.63\\
\botrule
\end{tabular*}
\begin{tablenotes}%
\item[$^{1}$] Mode of the posterior distribution of the parameter.
\item[$^{2}$] Equal-tailed 95\% credible intervals on the univariate posterior distributions of the parameters. \vspace*{6pt}
\end{tablenotes}
\end{table*}

\begin{figure*}[!t]%
\centering
\includegraphics[width=0.70\linewidth]{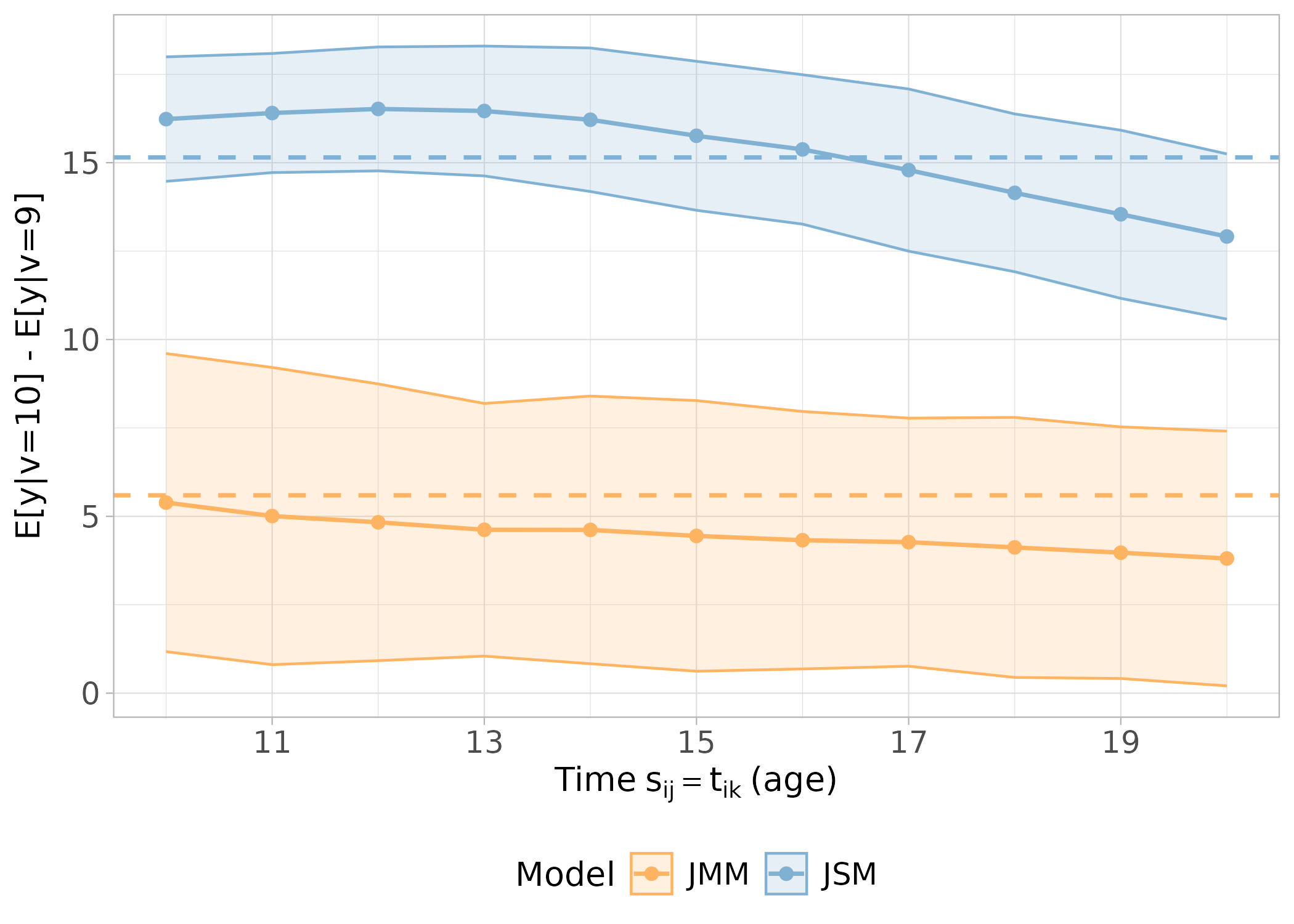}
\caption{Cross-sectional association between the protein and the PUL2.0 outcome across ages 10 to 20, with the base value $a$ set to 9. The comparison is between the Joint Mixed model (orange line) and the Joint Scaled Model (blue line). The over all time mean value of the association coefficient is also plotted (dashed lines).}\label{fig1}
\end{figure*}

Figure \ref{fig1} illustrates how the marginal association (\ref{beta joint}) between protein MYOM3 and the PUL2.0 outcome develops over time, assuming both variables are measured instantaneously, i.e., cross-sectional association. Note that the estimation of the integral (\ref{alpha}) requires the conditional mean of the outcome (\ref{cond mean}), which becomes a nonlinear function of the linear predictor when the outcome is non-normally distributed. As in the theory of GLMM \citep{diggle_analysis_2002, robert_e_weiss_modeling_2005}, this implies that the difference (\ref{beta joint}) is not constant but instead varies with the assumed value $a$ of the protein. For clarity of presentation, we initially report results on a reference value $a = 9$. This choice allows us to examine how the outcome changes when the predictor increases to its observed mean value, which is around 10 (Figure \ref{spaghetti plots}). We considered the age interval 10–20, as the PUL2.0 is typically not used in clinical practice to evaluate disease progression in younger patients.
The effect of the endogenous variable on the outcome estimated by the JMM does not change much throughout the observed period; a one-unit increase in the log-protein corresponds to around 5-point increase in PUL2.0 (Figure \ref{fig1}). 
In contrast, the JSM estimates a stronger association. It change slightly more over the considered time compared to the JMM, but on average, a one-unit increase in the log-protein results in a 14.62-point increase on the PUL2.0 test. 
The uncertainty around the JSM estimates is narrower than that for JMM.

\begin{figure}[t]
    \centering
    \begin{minipage}{0.49\textwidth}
        \centering
        \subfigure[Joint Mixed Model]{
            \includegraphics[width=\linewidth]{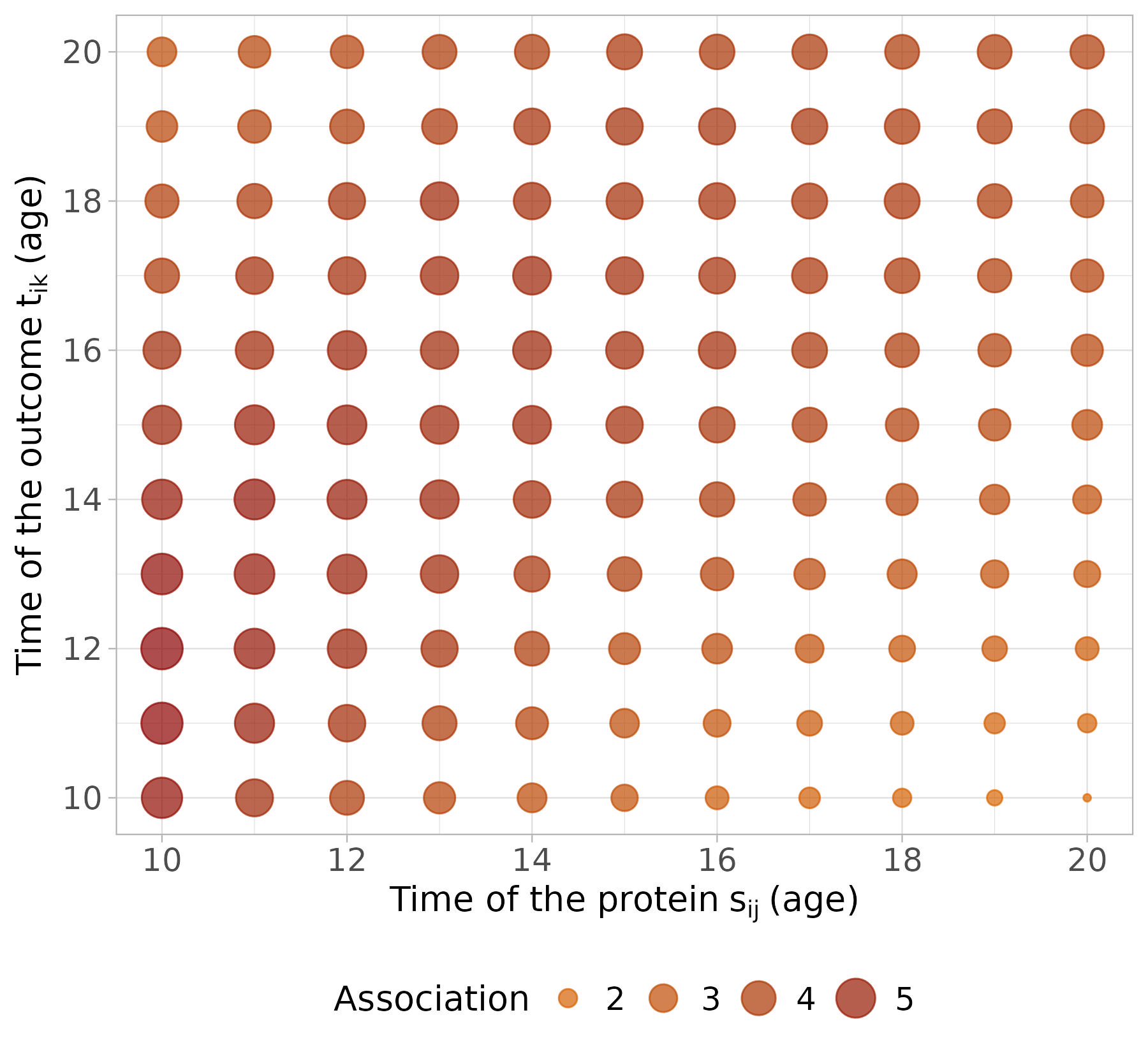}
        }
    \end{minipage}
    \hfill
    \begin{minipage}{0.49\textwidth}
        \centering
        \subfigure[Joint Scaled Model]{
            \includegraphics[width=\linewidth]{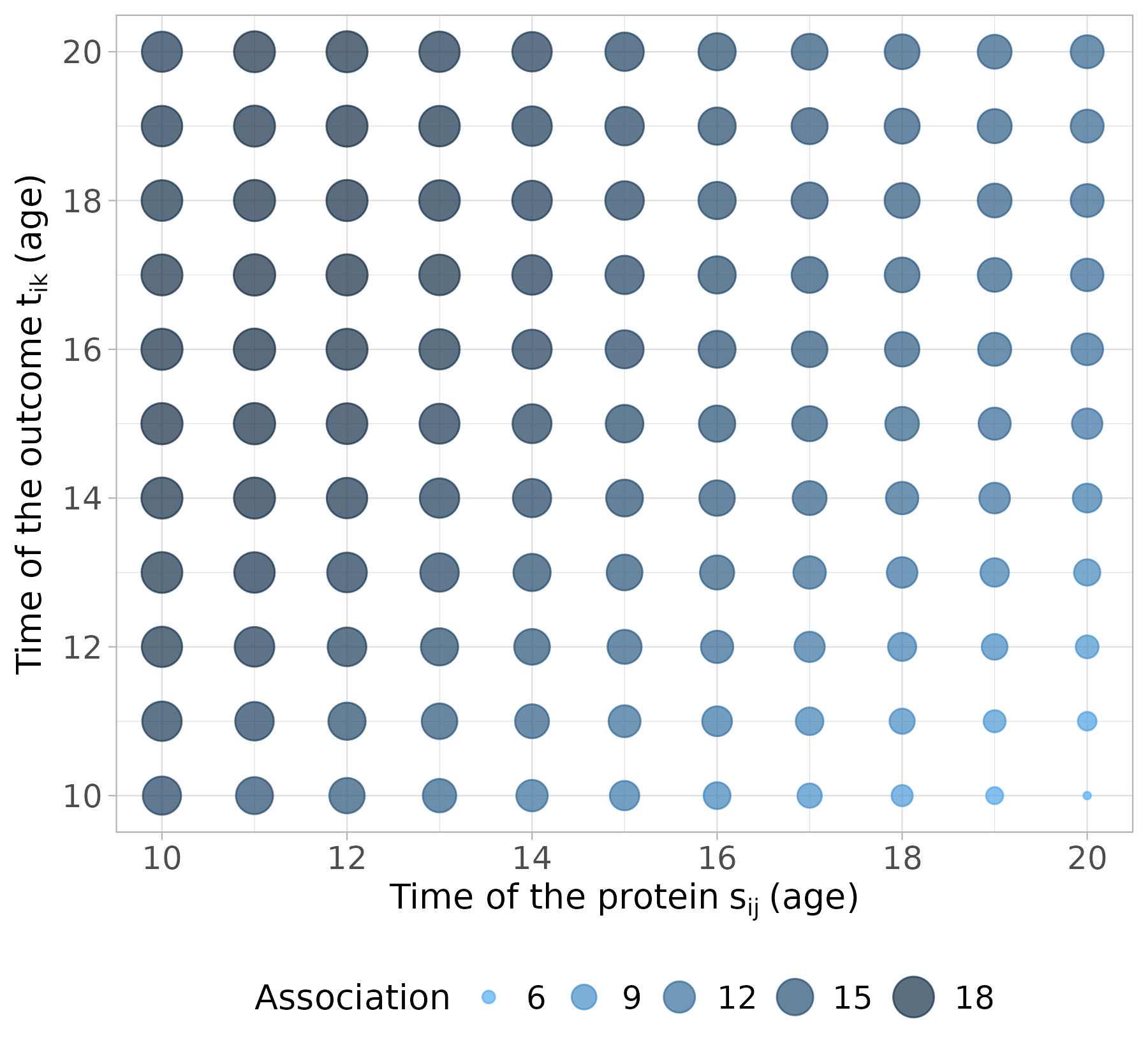}
        }
    \end{minipage}
    \caption{Lag association between the protein (x-axis) and the PUL2.0 outcome (y-axis) across ages 10 to 20, for the Joint Mixed Model and the Joint Scaled Model. Association is estimated assuming the endogenous variable equal to 9.}
    \label{fig2}
\end{figure}

A particularly insightful aspect of the analysis involves evaluating the lag effect of protein levels on the outcome. In biological and medical contexts, it is often of interest to understand how covariate values at previous time points influence future outcomes. 
Figure \ref{fig2} illustrates the lagged effects, where the lag is defined as the time distance between a protein change and the subsequent change in the response.
In both the JSM and JMM scenarios, we observe a strong association between past protein measurements and future outcome observations, as illustrated in the upper triangle of Figure \ref{fig2}. This association remains relatively stable even as the temporal gap between the two measurements increases. For example, in the JMM model, a unit increase in protein levels around age 10 corresponds to a 5-point increase in PUL2.0 at age 14, and a 4-point increase at age 18. Similarly, in the JSM model, the protein collected between ages 10 and 12 show the highest influence on outcomes measured around ages 15 to 18.
However, when the same lag distance is considered at later ages, the strength of the association tends to decline, likely due in part to reduced data availability from age 17. 
Overall, the temporal pattern of the lag effect is similar across both models, with the primary difference being the magnitude of the associations.

\begin{figure}[t]
    \centering
    \begin{minipage}{0.49\textwidth}
        \centering
        \subfigure[Joint Mixed Model]{
            \includegraphics[width=\linewidth]{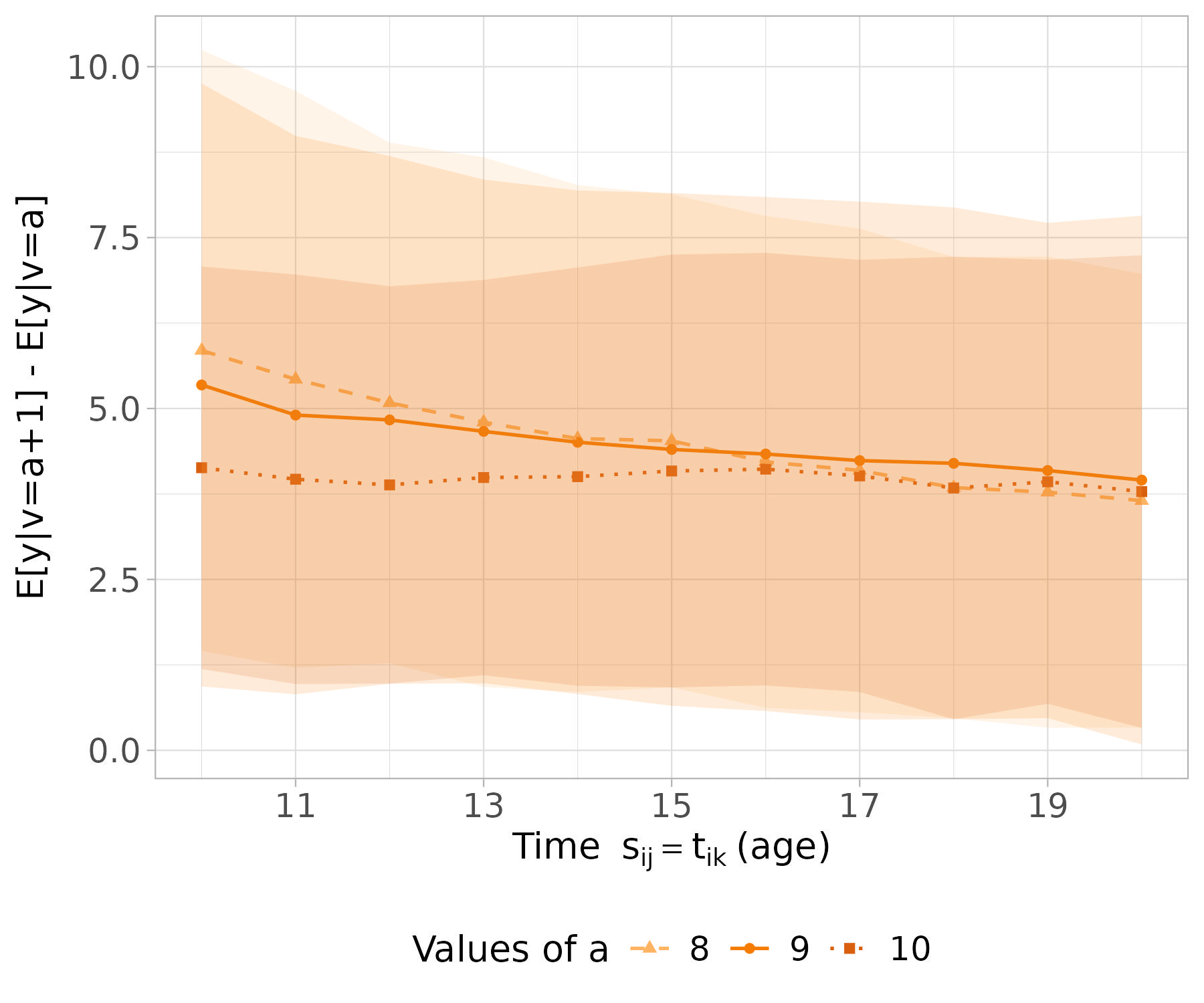}
        }
    \end{minipage}
    \hfill
    \begin{minipage}{0.49\textwidth}
        \centering
        \subfigure[Joint Scaled Model]{
            \includegraphics[width=\linewidth]{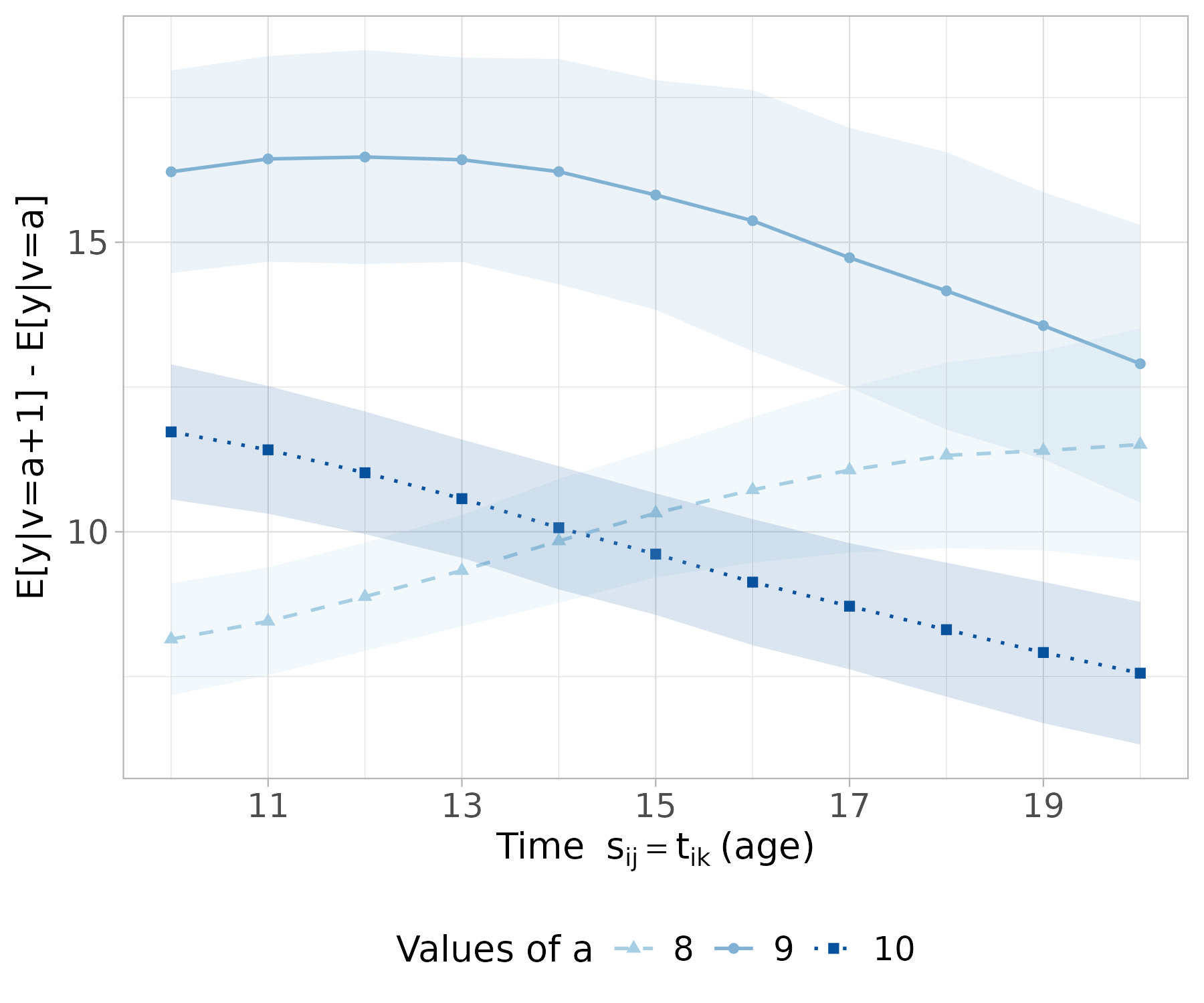}
        }
    \end{minipage}
    \caption{Cross-sectional association coefficient between the protein and the PUL2.0 outcome across ages 10 to 20, for different reference values of the proteins. Comparison between Joint Mixed Model and Joint Scaled Model.}
    \label{fig_all}
\end{figure}

Till now the results were presented assuming the value of the protein $a = 9$. Nevertheless, the association between the protein and the outcome can change based on the assumed values of the protein. In Figure \ref{fig_all} we present the estimation of the cross-sectional effect of the protein on the PUL2.0 outcome at different values of the protein. To ensure coverage across the full range of protein values observed in the data (Figure \ref{spaghetti plots}), we included results for references $a =8$, $a =9$, and $a = 10$. In the case of the JMM, the estimated association coefficients (\ref{beta joint}) remain relatively consistent across the three assumed scenarios. 
Regardless of the value of $a$, the influence of the endogenous variable on the PUL2.0 outcome tends to slightly diminish and stabilize once patients reach approximately 15 to 20 years of age. Overall, changes in protein levels appear to increase the mean PUL2.0 score from 5 to 4 points.
In contrast, the JSM shows greater sensitivity to the assumed value of the covariate. When the endogenous variable is set to 8, it increases the PUL2.0 outcome by approximately 8 units at younger ages, with the effect growing to around 12 units at older ages. The strongest association is observed when both the endogenous variable and the outcome are measured at around age 12, and the endogenous variable is set to 9. Conversely, the weakest effect occurs when the covariate is fixed at 10 and measured at age 20, resulting in an increase of only 5 units in the PUL2.0 outcome (Figure \ref{fig_all}).

The choice of the best model can be guided by several criteria: marginal likelihoods, the information criteria DIC and WAIC (Section \ref{INLA_estimation}). Results are shown in Table \ref{tab2}. Based on the marginal likelihood, the JMM appears to slightly outperform the JSM in describing the observed data. 
According to DIC and WAIC, however, the JSM offers a better balance between model fit and complexity than the JMM. Conclusions among the overall measures and the outcome measures are the same. 

\begin{table}[ht]
\caption{Evaluation of the models' performance based on Bayesian evaluation criteria. INLA provides two options for each measure: the "overall" option, which assesses model fit across all observations (both endogenous and response variables), and the "outcome" option, which evaluates how good is the model on predicting and fitting only the outcome part of the model, given the observed endogenous variable and estimated parameters. \label{tab2}}
\centering
\begin{tabular}{l@{\hskip 20pt}c@{\hskip 20pt}cc@{\hskip 20pt}cc@{\hskip 20pt}cc}
  \hline
 Model & \shortstack{Marginal\\Likelihood} & \shortstack{Overall\\DIC} & \shortstack{Outcome \\DIC} & \shortstack{Overall \\WAIC} & \shortstack{Outcome \\WAIC} 
 \\ 
  \hline
 JMM & 129.91 & -752.43 & -811.18 & -756.19 & -817.54 
 \\ 
 JSM & 121.47 & -782.07 & -825.66 & -786.61 & -831.44
 \\ 
   \hline
\end{tabular}
\end{table}

\begin{figure*}[!t]%
\centering
\includegraphics[width=0.95\linewidth]{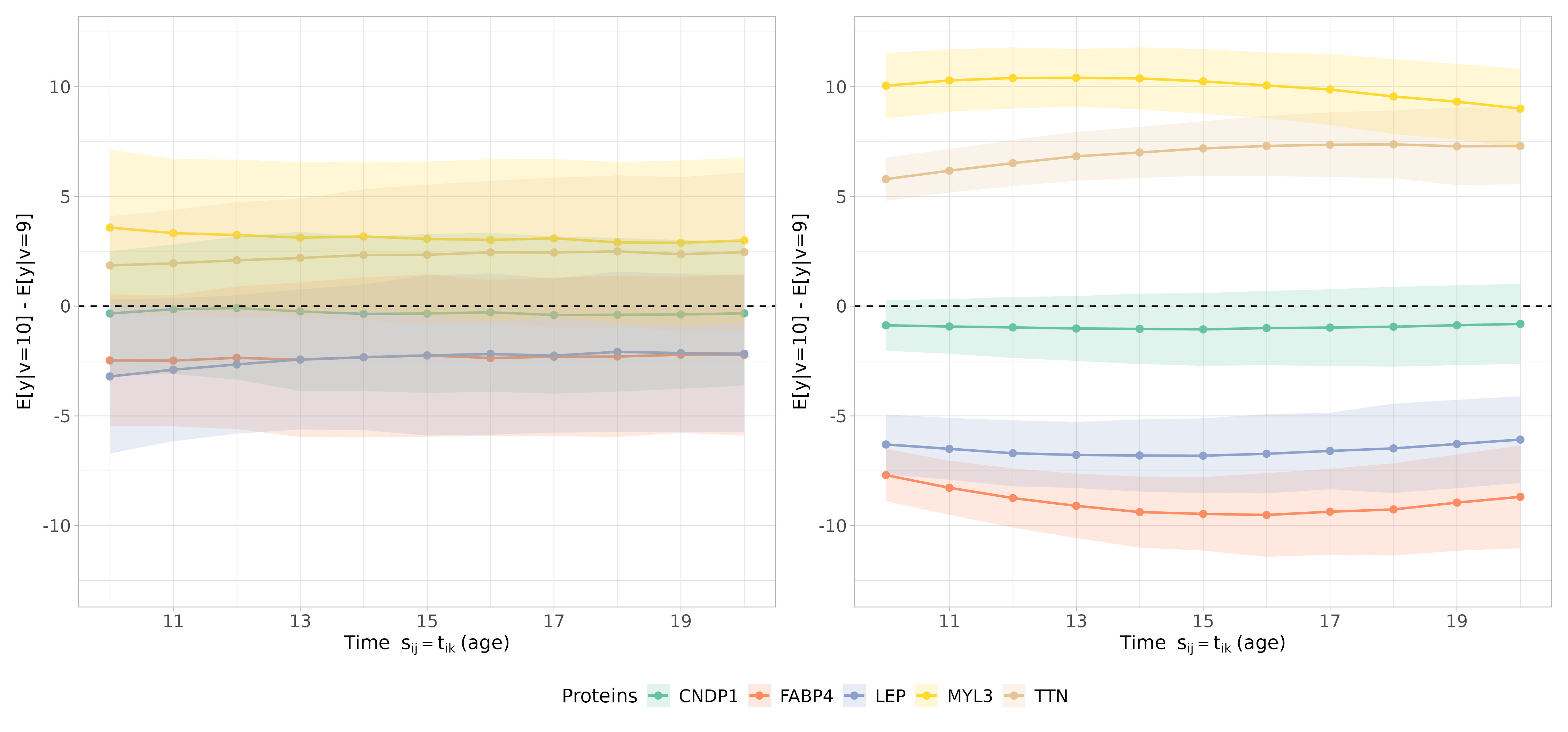}
\caption{Cross-sectional association between each protein and the PUL2.0 outcome across ages 10 to 20, when the endogenous variable is fixed equal to 9. The comparison is between the Joint Mixed Model (left panel) and the Joint Scaled Model (right panel). }
\label{more_prot}
\end{figure*}

Finally, we conducted the same analysis separately for five additional proteins (Figure \ref{more_prot}). While the direction of the estimated effects is concordant between JMM and JSM, their magnitudes differ. Both models exhibit relatively stable associations over time for each protein; however, in the JMM, the effect sizes remain within a narrow range (approximately between -5 and 5), indicating limited variability across proteins. In contrast, the JSM model shows a broader range of effect sizes (approximately between -10 and 10), suggesting greater sensitivity to protein-specific dynamics. Notably, both models indicate no association between CNDP1 and the PUL2.0 outcome.

\section{Discussion}\label{sec5}
In biomedical research, longitudinal outcomes and endogenous time-dependent covariates (e.g., biomarkers) are often recorded, creating the need to develop methodological approaches to assess their associations, evaluate how one outcome changes with the covariate, and determine how this relationship evolves over time. To address these aspects, endogenous covariate and outcome are typically modeled jointly by assuming correlated random effects. We refered to this model as the Joint Mixed Model (JMM). 
An alternative approach is a modification of the joint model proposed by \cite{rizopoulos_joint_2012}, adapting it to include only longitudinal outcomes rather than a time-to-event component. We refered to this adapted model as the Joint Scaled Model (JSM). 
However, making inference about unknown association of the endogenous time-varying variable on the outcome is not straightforward in either model. To address these limitation, we propose to numerically derive an association coefficient (\ref{alpha}) that offers several advantages:
\begin{itemize}
    \item Quantifies marginal effects. Unlike the JMM, where the association is inferred through correlations between random effects (latent space), and the JSM, where the effect is conditional on random effects (subject-specific effect) (Table \ref{tab1}), our method estimates the expected change in the outcome for a unit change in the covariate. This population-level estimate is more informative for clinical decision-making, trial endpoints definition, and biomarker validation; 
    \item Improves interpretability. It reflects changes in the actual scale of the outcome, even when the outcome is non-normally distributed. In contrast, JMM and JSM expresses effects on the transformed mean (e.g., logit or probit), which can be difficult to communicate in applied settings. 
    \item Enables the comparison of the two model results and assesses the impact of their underlying assumptions on conclusions in the motivating example (Section \ref{sec4}). 
    \item Captures dynamic association. It allows the strength of the association to vary over time, reflecting how the relationship between the covariate and the outcome evolves (Figure \ref{fig1}, Figure \ref{more_prot}) with patient age or disease progression. This is particularly valuable in progressive disorders such as Duchenne Muscular Dystrophy, where biomarker relevance may change across disease stages.
    \item Assesses lagged affect. It enables evaluation of how past values of the covariate influence future outcomes, offering insights into delayed effects (Figure \ref{fig2}). In many biological systems, changes in a biomarker may not have an immediate effect on the outcome, so being able to assess lagged associations is essential for understanding disease mechanisms.
    \item Examines how the relationship between the covariate and the outcome changes when conditioning on different values of the endogenous variable. While the average effect can also be considered, this may oversimplify the analysis, especially when the conditional mean $E[y_{ik}|v_{ij}]$ is non-linear.
\end{itemize}

Note that in this paper we focused on a setting with only a single continuous endogenous variable and one response that can be of any type. A limitation of the association coefficient (\ref{alpha}) that we derive is that it no longer applies when non-normal endogenous variables are considered. The normality assumption of the endogenous variable ensures that the posterior distribution of the random effects remains Gaussian, allowing its form to be defined analytically. Once this assumption no longer holds, distributions (\ref{cond mean}) and (\ref{cond b}), required for estimating the integral (\ref{alpha}), can no longer be analytically expressed. Consequently, alternative numerical approximation methods must be employed to estimate the integral. 

Some considerations can be made regarding the structural assumptions and flexibility of the two models. In the JMM, by assuming a form of the variance-covariance matrix of random effects, we impose some constraints on the form of association of variables and limit the flexibility of the association. 
The JSM model, instead, can offer greater flexibility; e.g., by allowing distinct scaling parameters for the components of the endogenous variable's linear predictor that are copied and scaled within the outcome's linear predictor, or by relaxing the assumption of uncorrelated random effects and incorporating correlation at the latent level as well.
However, further studies need to be done to verify the actual flexibility of the JSM, particularly with regard to the number of parameters that can be reliably estimated before the model becomes unstable. 

In this work, we primarily assessed the stability of the two joint models under varying sample sizes (Section \ref{simulation}) and examined how their performance is affected by the choice of hyperprior distributions (Section 3.1 of the \href{Supplementary_material/main.tex}{Supplementary Material}). Simulation showed that overall the JSM tends to be less sensitive than the JMM (Figure \ref{figsim}) to the amount of observations available. In other words, the JSM seems to be more stable that the JMM. However, while the JSM tends to consistently overestimate the association coefficient $\beta_{joint}$ at both $N = 65$ and $N = 200$, the JMM produces more accurate estimates when the sample size is sufficiently large.
The choice of hyperpriors has a notable impact on the estimation accuracy of the hyperparameters for both models. Using informative priors substantially improves performance, not only in small-sample settings ($N = 65$) but also when larger samples are available ($N = 200$). Nonetheless, Figures~\ref{fig_sim_jmm} and~\ref{fig_sim_jsm} show that, although informative hyperpriors improve estimation, further gains may still be possible through more careful prior specification or alternative prior choices. We recommend that future applications of both JMM and JSM pay careful attention to prior specification, as it can significantly influence the reliability of the resulting association estimates. In future applications where both models are considered, it is recommended to assess their performance using information-based criteria, as shown in Section \ref{sec4}. Such evaluation can help in identifying the model most suitable for the data.

Finally, we implemented the models using the INLA framework, which has previously demonstrated efficient estimation in settings involving multiple likelihoods \citep{martins_bayesian_2013}. Our results further confirm that INLA can be effectively applied to estimate both models considered in this study. One limitation of INLA in the context of our project is that it provides only marginal posterior distributions, not conditional ones. As a result, estimating the association coefficient required additional computational effort through Monte Carlo methods, increasing the overall computational burden. A second limitation lies in the complexity of implementing joint models within the INLA framework. This process required a substantial reformulation of the data structure, particularly for defining the JSM. While the R package INLAjoint \citep{rustand_fast_2024} could have facilitated the estimation process, it was limited to the JMM and did not provide access to the full range of features available in the INLA framework. In particular, the package lacked flexibility for tasks such as specifying custom hyperpriors. To ensure greater adaptability across modeling scenarios, we ultimately developed our own implementation using INLA directly (https://github.com/cdegan/Multivariate-longitudinal-modeling-with-endogenous-variables/tree/main).

\section{Supplementary material}
The supplementary material is available online. The real-word dataset used as motivating study is not available, but anonymized data can be made available to qualified investigators on request. Requests should be in line with the approved ethical protocol.

\section{Software}
The code of the simulation and of the implementation of the two models is available in https://github.com/cdegan/Multivariate-longitudinal-modeling-with-endogenous-variables/tree/main. 

\section{Funding}
This work was supported by the National Institute Of Neurological Disorders And Stroke of the National Institutes of Health (NIH) under Award Number [R61NS119639] (Co-I: Spitali, Tsonaka). The collection of the data used as motivating study was funded by Parent Project Muscular Dystrophy through the Protein Mapping Project.

\section{Acknowledgments}
Several authors of this publication are members of the Netherlands Neuromuscular Center (NL-NMD) and the European Reference Network for rare neuromuscular diseases EURO-NMD. The authors thank all Duchenne Muscular Dystrophy patients and their families for contributing to the study. The authors also thank Georgy Gomon, whose master's thesis provides valuable insights into the coding of these two models in INLA.

\section{Conflict interests}
The authors have no competing interests.

\bibliographystyle{abbrvnat}
\bibliography{reference3}

@article{qian_linear_2020,
	title = {Linear {Mixed} {Models} with {Endogenous} {Covariates}: {Modeling} {Sequential} {Treatment} {Effects} with {Application} to a {Mobile} {Health} {Study}},
	volume = {35},
	issn = {0883-4237},
	shorttitle = {Linear {Mixed} {Models} with {Endogenous} {Covariates}},
	url = {https://www.jstor.org/stable/26997907},
	number = {3},
	urldate = {2024-06-18},
	journal = {Statistical Science},
	author = {Qian, Tianchen and Klasnja, Predrag and Murphy, Susan A.},
	year = {2020},
	note = {Publisher: Institute of Mathematical Statistics},
	pages = {375--390},
}

@book{robert_e_weiss_modeling_2005,
	address = {New York, NY},
	series = {Springer {Texts} in {Statistics}},
	title = {Modeling {Longitudinal} {Data}},
	copyright = {http://www.springer.com/tdm},
	isbn = {978-0-387-40271-0},
	url = {http://link.springer.com/10.1007/0-387-28314-5},
	language = {en},
	urldate = {2024-06-21},
	publisher = {Springer},
	author = {Robert E. Weiss},
	year = {2005},
	doi = {10.1007/0-387-28314-5},
}

@book{wu_mixed_2009,
	address = {New York},
	title = {Mixed {Effects} {Models} for {Complex} {Data}},
	isbn = {978-0-429-14251-2},
	publisher = {Chapman and Hall/CRC},
	author = {Wu, Lang},
	month = nov,
	year = {2009},
	doi = {10.1201/9781420074086},
}

@book{rizopoulos_joint_2012,
	address = {New York},
	title = {Joint {Models} for {Longitudinal} and {Time}-to-{Event} {Data}: {With} {Applications} in {R}},
	isbn = {978-0-429-06338-1},
	shorttitle = {Joint {Models} for {Longitudinal} and {Time}-to-{Event} {Data}},
	publisher = {Chapman and Hall/CRC},
	author = {Rizopoulos, Dimitris},
	month = jun,
	year = {2012},
	doi = {10.1201/b12208},
}

@article{verbeke_analysis_2012,
	title = {The analysis of multivariate longitudinal data: {A} review},
	volume = {23},
	issn = {0962-2802},
	shorttitle = {The analysis of multivariate longitudinal data},
	url = {https://www.ncbi.nlm.nih.gov/pmc/articles/PMC3404254/},
	doi = {10.1177/0962280212445834},
	number = {1},
	urldate = {2025-01-29},
	journal = {Statistical Methods in Medical Research},
	author = {Verbeke, Geert and Fieuws, Steffen and Molenberghs, Geert and Davidian, Marie},
	year = {2012},
	pmid = {22523185},
	pmcid = {PMC3404254},
	pages = {42--59},
}

@book{fitzmaurice_longitudinal_2008,
	address = {New York},
	title = {Longitudinal {Data} {Analysis}},
	isbn = {978-0-429-14267-3},
	publisher = {Chapman and Hall/CRC},
	editor = {Fitzmaurice, Garrett and Davidian, Marie and Verbeke, Geert and Molenberghs, Geert},
	month = aug,
	year = {2008},
	doi = {10.1201/9781420011579},
}

@book{molenberghs_models_2005,
	address = {New York},
	series = {Springer {Series} in {Statistics}},
	title = {Models for {Discrete} {Longitudinal} {Data}},
	copyright = {http://www.springer.com/tdm},
	isbn = {978-0-387-25144-8},
	url = {http://link.springer.com/10.1007/0-387-28980-1},
	language = {en},
	urldate = {2025-01-30},
	publisher = {Springer-Verlag},
	author = {Molenberghs, Geert and Verbeke, Geert},
	year = {2005},
	doi = {10.1007/0-387-28980-1},
}

@article{mcculloch_joint_2016,
	title = {Joint modelling of mixed outcome types using latent variables},
	volume = {17},
	issn = {0962-2802},
	doi = {10.1177/0962280207081240},
	number = {1},
	journal = {Statistical Methods in Medical Research},
	author = {McCulloch, Charles},
	year = {2016},
	pmid = {17855745},
	pages = {53--73},
}

@article{rustand_fast_2024,
	title = {Fast and flexible inference for joint models of multivariate longitudinal and survival data using integrated nested {Laplace} approximations},
	volume = {25},
	issn = {1465-4644},
	url = {https://doi.org/10.1093/biostatistics/kxad019},
	doi = {10.1093/biostatistics/kxad019},
	number = {2},
	urldate = {2025-02-05},
	journal = {Biostatistics},
	author = {Rustand, Denis and van Niekerk, Janet and Krainski, Elias Teixeira and Rue, Håvard and Proust-Lima, Cécile},
	month = apr,
	year = {2024},
	pages = {429--448},
}

@book{little_statistical_2019,
	edition = {1},
	series = {Wiley {Series} in {Probability} and {Statistics}},
	title = {Statistical {Analysis} with {Missing} {Data}, {Third} {Edition}},
	copyright = {http://doi.wiley.com/10.1002/tdm\_license\_1.1},
	isbn = {978-0-470-52679-8 978-1-119-48226-0},
	url = {https://onlinelibrary.wiley.com/doi/book/10.1002/9781119482260},
	language = {en},
	urldate = {2025-03-05},
	publisher = {Wiley},
	author = {Little, Roderick and Rubin, Donald},
	month = apr,
	year = {2019},
	doi = {10.1002/9781119482260},
}

@article{iddi_joint_2012,
	title = {A joint marginalized multilevel model for longitudinal outcomes},
	volume = {39},
	issn = {0266-4763},
	url = {https://doi.org/10.1080/02664763.2012.711302},
	doi = {10.1080/02664763.2012.711302},
	number = {11},
	urldate = {2025-03-05},
	journal = {Journal of Applied Statistics},
	author = {Iddi, Samuel and Molenberghs, Geert},
	month = nov,
	year = {2012},
	note = {Publisher: Taylor \& Francis
\_eprint: https://doi.org/10.1080/02664763.2012.711302},
	pages = {2413--2430},
}

@article{ivanova_mixed_2016,
	title = {Mixed models approaches for joint modeling of different types of responses},
	volume = {26},
	issn = {1054-3406},
	url = {https://doi.org/10.1080/10543406.2015.1052487},
	doi = {10.1080/10543406.2015.1052487},
	number = {4},
	urldate = {2025-03-05},
	journal = {Journal of Biopharmaceutical Statistics},
	author = {Ivanova, Anna and Molenberghs, Geert and Verbeke, Geert},
	month = jul,
	year = {2016},
	pmid = {26098411},
	note = {Publisher: Taylor \& Francis
\_eprint: https://doi.org/10.1080/10543406.2015.1052487},
	pages = {601--618},
}

@article{tsiatis_joint_2004,
	title = {Joint {Modeling} of {Longitudinal} and {Time}-to-{Event} {Data}: {An} {Overview}},
	volume = {14},
	issn = {1017-0405},
	shorttitle = {Joint {Modeling} of {Longitudinal} and {Time}-to-{Event} {Data}},
	url = {https://www.jstor.org/stable/24307417},
	number = {3},
	urldate = {2025-03-05},
	journal = {Statistica Sinica},
	author = {Tsiatis, Anastasios A. and Davidian, Marie},
	year = {2004},
	note = {Publisher: Institute of Statistical Science, Academia Sinica},
	pages = {809--834},
}

@article{geman_stochastic_1984,
	title = {Stochastic {Relaxation}, {Gibbs} {Distributions}, and the {Bayesian} {Restoration} of {Images}},
	volume = {PAMI-6},
	issn = {1939-3539},
	url = {https://ieeexplore.ieee.org/document/4767596},
	doi = {10.1109/TPAMI.1984.4767596},
	number = {6},
	urldate = {2025-03-05},
	journal = {IEEE Transactions on Pattern Analysis and Machine Intelligence},
	author = {Geman, Stuart and Geman, Donald},
	month = nov,
	year = {1984},
	note = {Conference Name: IEEE Transactions on Pattern Analysis and Machine Intelligence},
	pages = {721--741},
}

@article{metropolis_equation_1953,
	title = {Equation of {State} {Calculations} by {Fast} {Computing} {Machines}},
	volume = {21},
	issn = {0021-9606},
	url = {https://doi.org/10.1063/1.1699114},
	doi = {10.1063/1.1699114},
	number = {6},
	urldate = {2025-03-05},
	journal = {The Journal of Chemical Physics},
	author = {Metropolis, Nicholas and Rosenbluth, Arianna W. and Rosenbluth, Marshall N. and Teller, Augusta H. and Teller, Edward},
	month = jun,
	year = {1953},
	pages = {1087--1092},
}

@book{gomez-rubio_bayesian_2021,
	title = {Bayesian inference with {INLA}},
	url = {http://becarioprecario.bitbucket.io/inla-gitbook/index.html},
	urldate = {2025-03-06},
	publisher = {Chapman and Hall/CRC},
	author = {Gómez-Rubio, Virgilio},
	year = {2021},
}

@article{niekerk_new_2021,
	title = {New {Frontiers} in {Bayesian} {Modeling} {Using} the {INLA} {Package} in {R}},
	volume = {100},
	copyright = {Copyright (c) 2021 Janet van Niekerk, Haakon Bakka, Håvard Rue, Olaf Schenk},
	issn = {1548-7660},
	url = {https://doi.org/10.18637/jss.v100.i02},
	doi = {10.18637/jss.v100.i02},
	language = {en},
	number = {2},
	urldate = {2025-03-10},
	journal = {Journal of Statistical Software},
	author = {Niekerk, Janet Van and Bakka, Haakon and Rue, Håvard and Schenk, Olaf},
	month = nov,
	year = {2021},
	pages = {1--28},
}

@article{bonat_likelihood_2014,
	title = {Likelihood analysis for a class of beta mixed models},
	volume = {42},
	issn = {0266-4763},
	url = {https://doi.org/10.1080/02664763.2014.947248},
	doi = {10.1080/02664763.2014.947248},
	number = {2},
	urldate = {2025-03-10},
	journal = {Journal of Applied Statistics},
	author = {Bonat, Wagner Hugo and Ribeiro Jr, Paulo Justiniano and Zeviani, Walmes Marques},
	year = {2014},
	note = {Publisher: Taylor \& Francis
\_eprint: https://doi.org/10.1080/02664763.2014.947248},
	pages = {252--266},
}

@article{figueroa-zuniga_mixed_2013,
	title = {Mixed beta regression: {A} {Bayesian} perspective},
	volume = {61},
	issn = {0167-9473},
	shorttitle = {Mixed beta regression},
	url = {https://www.sciencedirect.com/science/article/pii/S0167947312004239},
	doi = {10.1016/j.csda.2012.12.002},
	urldate = {2025-03-10},
	journal = {Computational Statistics \& Data Analysis},
	author = {Figueroa-Zúñiga, Jorge I. and Arellano-Valle, Reinaldo B. and Ferrari, Silvia L. P.},
	month = may,
	year = {2013},
	pages = {137--147},
}

@article{mayhew_performance_2020,
	title = {Performance of {Upper} {Limb} module for {Duchenne} muscular dystrophy},
	volume = {62},
	copyright = {© 2019 Mac Keith Press},
	issn = {1469-8749},
	url = {https://onlinelibrary.wiley.com/doi/abs/10.1111/dmcn.14361},
	doi = {10.1111/dmcn.14361},
	language = {en},
	number = {5},
	urldate = {2025-03-10},
	journal = {Developmental Medicine \& Child Neurology},
	author = {Mayhew, Anna G and Coratti, Giorgia and Mazzone, Elena Stacy and Klingels, Katrijn and James, Meredith and Pane, Marika and Straub, Volker and Goemans, Natalie and Mercuri, Eugenio and Group, Pul Working},
	year = {2020},
	note = {\_eprint: https://onlinelibrary.wiley.com/doi/pdf/10.1111/dmcn.14361},
	pages = {633--639},
}

@article{delporte_joint_2022,
	title = {A joint normal-binary (probit) model},
	volume = {90},
	copyright = {© 2022 International Statistical Institute.},
	issn = {1751-5823},
	url = {https://onlinelibrary.wiley.com/doi/abs/10.1111/insr.12532},
	doi = {10.1111/insr.12532},
	language = {en},
	number = {S1},
	urldate = {2025-03-10},
	journal = {International Statistical Review},
	author = {Delporte, Margaux and Fieuws, Steffen and Molenberghs, Geert and Verbeke, Geert and Situma Wanyama, Simeon and Hatziagorou, Elpis and De Boeck, Christiane},
	year = {2022},
	note = {\_eprint: https://onlinelibrary.wiley.com/doi/pdf/10.1111/insr.12532},
	pages = {S37--S51},
}

@article{martins_bayesian_2013,
	title = {Bayesian computing with {INLA}: {New} features},
	volume = {67},
	issn = {0167-9473},
	shorttitle = {Bayesian computing with {INLA}},
	url = {https://www.sciencedirect.com/science/article/pii/S0167947313001552},
	doi = {10.1016/j.csda.2013.04.014},
	urldate = {2025-03-10},
	journal = {Computational Statistics \& Data Analysis},
	author = {Martins, Thiago G. and Simpson, Daniel and Lindgren, Finn and Rue, Håvard},
	month = nov,
	year = {2013},
	pages = {68--83},
}

@article{fisher_time-dependent_1999,
	title = {{TIME}-{DEPENDENT} {COVARIATES} {IN} {THE} {COX} {PROPORTIONAL}-{HAZARDS} {REGRESSION} {MODEL}},
	volume = {20},
	issn = {0163-7525, 1545-2093},
	url = {https://www.annualreviews.org/content/journals/10.1146/annurev.publhealth.20.1.145},
	doi = {10.1146/annurev.publhealth.20.1.145},
	language = {en},
	number = {Volume 20, 1999},
	urldate = {2025-03-25},
	journal = {Annual Review of Public Health},
	author = {Fisher, Lloyd D. and Lin, D. Y.},
	month = may,
	year = {1999},
	note = {Publisher: Annual Reviews},
	pages = {145--157},
}

@article{pepe_modeling_1997,
	title = {Modeling {Partly} {Conditional} {Means} with {Longitudinal} {Data}},
	volume = {92},
	issn = {0162-1459},
	url = {https://doi.org/10.1080/01621459.1997.10474054},
	doi = {10.1080/01621459.1997.10474054},
	number = {439},
	urldate = {2025-04-08},
	journal = {Journal of the American Statistical Association},
	author = {Pepe, Margaret Sullivan and Couper, David},
	month = sep,
	year = {1997},
	note = {Publisher: ASA Website
\_eprint: https://doi.org/10.1080/01621459.1997.10474054},
	pages = {991--998},
}

@article{pinheiro_approximations_1995,
	title = {Approximations to the {Log}-{Likelihood} {Function} in the {Nonlinear} {Mixed}-{Effects} {Model}},
	volume = {4},
	issn = {1061-8600},
	url = {https://www.tandfonline.com/doi/abs/10.1080/10618600.1995.10474663},
	doi = {10.1080/10618600.1995.10474663},
	number = {1},
	urldate = {2025-04-23},
	journal = {Journal of Computational and Graphical Statistics},
	author = {Pinheiro, José C. and and Bates, Douglas M.},
	month = mar,
	year = {1995},
	note = {Publisher: ASA Website
\_eprint: https://www.tandfonline.com/doi/pdf/10.1080/10618600.1995.10474663},
	pages = {12--35},
}

@article{drikvandi_framework_2024,
	title = {A framework for analysing longitudinal data involving time-varying covariates},
	volume = {18},
	issn = {1932-6157, 1941-7330},
	doi = {10.1214/23-AOAS1851},
	number = {2},
	urldate = {2025-04-23},
	journal = {The Annals of Applied Statistics},
	author = {Drikvandi, Reza and Verbeke, Geert and Molenberghs, Geert},
	month = jun,
	year = {2024},
	note = {Publisher: Institute of Mathematical Statistics},
	pages = {1618--1641},
}

@article{seyoum_joint_2017,
	title = {Joint longitudinal data analysis in detecting determinants of {CD4} cell count change and adherence to highly active antiretroviral therapy at {Felege} {Hiwot} {Teaching} and {Specialized} {Hospital}, {North}-west {Ethiopia} ({Amhara} {Region})},
	volume = {14},
	issn = {1742-6405},
	url = {https://doi.org/10.1186/s12981-017-0141-3},
	doi = {10.1186/s12981-017-0141-3},
	number = {1},
	urldate = {2025-04-23},
	journal = {AIDS Research and Therapy},
	author = {Seyoum, Awoke and Ndlovu, Principal and Temesgen, Zewotir},
	month = mar,
	year = {2017},
	pages = {14},
}

@article{kassahun_joint_2013,
	title = {A joint model for hierarchical continuous and zero-inflated overdispersed count data},
	volume = {85},
	copyright = {© 2013 Taylor \& Francis},
	issn = {0094-9655},
	url = {https://www.tandfonline.com/doi/abs/10.1080/00949655.2013.829058},
	doi = {https://doi.org/10.1080/00949655.2013.829058},
	number = {3},
	urldate = {2025-04-23},
	journal = {Journal of Statistical Computation and Simulation},
	author = {Kassahun, Wondwosen and Neyens, Thomas and Molenberghs, Geert and Faes, Christel and Verbeke, Geert},
	year = {2013},
	note = {Publisher: Taylor \& Francis},
	pages = {552--571},
}

@article{amini_longitudinal_2021,
	title = {Longitudinal {Joint} {Modelling} of {Ordinal} and {Overdispersed} {Count} {Outcomes}: {A} {Bridge} {Distribution} for the {Ordinal} {Random} {Intercept}},
	copyright = {Copyright © 2021 Payam Amini et al.},
	issn = {1748-6718},
	shorttitle = {Longitudinal {Joint} {Modelling} of {Ordinal} and {Overdispersed} {Count} {Outcomes}},
	url = {https://onlinelibrary.wiley.com/doi/abs/10.1155/2021/5521881},
	doi = {10.1155/2021/5521881},
	urldate = {2025-04-23},
	journal = {Computational and Mathematical Methods in Medicine},
	author = {Amini, Payam and Moghimbeigi, Abbas and Zayeri, Farid and Tapak, Leili and Maroufizadeh, Saman and Verbeke, Geert},
	year = {2021},
	note = {\_eprint: https://onlinelibrary.wiley.com/doi/pdf/10.1155/2021/5521881},
}

@article{amini_longitudinal_2018,
	title = {Longitudinal {Joint} {Modelling} of {Binary} and {Continuous} {Outcomes}: {A} {Comparison} of {Bridge} and {Normal} {Distributions}},
	volume = {15},
	copyright = {Copyright (c) 2022},
	issn = {2282-0930},
	shorttitle = {Longitudinal {Joint} {Modelling} of {Binary} and {Continuous} {Outcomes}},
	url = {https://riviste.unimi.it/index.php/ebph/article/view/17434},
	doi = {10.2427/12755},
	language = {en},
	number = {1},
	urldate = {2025-04-23},
	journal = {Epidemiology, Biostatistics, and Public Health},
	author = {Amini, Payam and Verbeke, Geert and Zayeri, Farid and Mahjub, Hossein and Maroufizadeh, Saman and Moghimbeigi, Abbas},
	year = {2018},
	note = {Number: 1},
}

@article{efendi_joint_2013,
	title = {A joint model for longitudinal continuous and time-to-event outcomes with direct marginal interpretation},
	volume = {55},
	copyright = {© 2013 WILEY-VCH Verlag GmbH \& Co. KGaA, Weinheim},
	issn = {1521-4036},
	url = {https://onlinelibrary.wiley.com/doi/abs/10.1002/bimj.201200159},
	doi = {10.1002/bimj.201200159},
	language = {en},
	number = {4},
	urldate = {2025-04-24},
	journal = {Biometrical Journal},
	author = {Efendi, Achmad and Molenberghs, Geert and Njagi, Edmund Njeru and Dendale, Paul},
	year = {2013},
	note = {\_eprint: https://onlinelibrary.wiley.com/doi/pdf/10.1002/bimj.201200159},
	pages = {572--588},
}

@article{njagi_flexible_2013,
	title = {A flexible joint modeling framework for longitudinal and time-to-event data with overdispersion},
	volume = {25},
	issn = {0962-2802},
	url = {https://doi.org/10.1177/0962280213495994},
	doi = {10.1177/0962280213495994},
	number = {4},
	urldate = {2025-04-24},
	journal = {Statistical Methods in Medical Research},
	author = {Njagi, Edmund N and Molenberghs, Geert and Rizopoulos, Dimitris and Verbeke, Geert and Kenward, Michael G and Dendale, Paul and Willekens, Koen},
	year = {2013},
	note = {Publisher: SAGE Publications Ltd STM},
	pages = {1661--1676},
}

@article{gelfand_sampling-based_1990,
	title = {Sampling-{Based} {Approaches} to {Calculating} {Marginal} {Densities}},
	volume = {85},
	issn = {0162-1459},
	url = {https://www.jstor.org/stable/2289776},
	doi = {10.2307/2289776},
	number = {410},
	urldate = {2025-05-01},
	journal = {Journal of the American Statistical Association},
	author = {Gelfand, Alan E. and Smith, Adrian F. M.},
	year = {1990},
	note = {Publisher: [American Statistical Association, Taylor \& Francis, Ltd.]},
	pages = {398--409},
}

@book{diggle_analysis_2002,
	title = {Analysis of {Longitudinal} {Data}},
	isbn = {978-0-19-852484-7},
	url = {https://doi.org/10.1093/oso/9780198524847.001.0001},
	urldate = {2025-09-26},
	publisher = {Oxford University Press},
	author = {Diggle, Peter J and Heagerty, Patrick J and Liang, Kung-yee. and Zeger, Scott L},
	month = jun,
	year = {2002},
	doi = {10.1093/oso/9780198524847.001.0001},
}

@article{duan_duchenne_2021,
	title = {Duchenne muscular dystrophy},
	volume = {7},
	copyright = {2021 Springer Nature Limited},
	issn = {2056-676X},
	url = {https://www.nature.com/articles/s41572-021-00248-3},
	doi = {10.1038/s41572-021-00248-3},
	language = {en},
	number = {1},
	urldate = {2025-09-26},
	journal = {Nature Reviews Disease Primers},
	author = {Duan, Dongsheng and Goemans, Nathalie and Takeda, Shin’ichi and Mercuri, Eugenio and Aartsma-Rus, Annemieke},
	month = feb,
	year = {2021},
	note = {Publisher: Nature Publishing Group},
	pages = {13},
}

@book{buuren_flexible_2018,
	address = {New York},
	edition = {2},
	title = {Flexible {Imputation} of {Missing} {Data}, {Second} {Edition}},
	isbn = {978-0-429-49225-9},
	publisher = {Chapman and Hall/CRC},
	author = {Buuren, Stef van},
	month = jul,
	year = {2018},
	doi = {10.1201/9780429492259},
}

@article{li_considering_2023,
	title = {Considering time-lag effects can improve the accuracy of {NPP} simulation using a light use efficiency model},
	volume = {33},
	issn = {1861-9568},
	url = {https://doi.org/10.1007/s11442-023-2115-9},
	doi = {10.1007/s11442-023-2115-9},
	number = {5},
	journal = {Journal of Geographical Sciences},
	author = {Li, Chuanhua and Liu, Yunfan and Zhu, Tongbin and Zhou, Min and Dou, Tianbao and Liu, Lihui and Wu, Xiaodong},
	month = may,
	year = {2023},
	pages = {961--979},
}

@article{delporte_joint_2025,
	title = {A joint normal-ordinal (probit) model for ordinal and continuous longitudinal data},
	volume = {26},
	issn = {1465-4644},
	url = {https://doi.org/10.1093/biostatistics/kxae014},
	doi = {10.1093/biostatistics/kxae014},
	number = {1},
	urldate = {2025-09-26},
	journal = {Biostatistics},
	author = {Delporte, Margaux and Molenberghs, Geert and Fieuws, Steffen and Verbeke, Geert},
	month = jan,
	year = {2025},
	pages = {kxae014},
}

@mastersthesis{gomon_georgy_joint_2022,
	title = {Joint {Models}: {Implementation} in {INLA} and {Applications}},
	school = {Leiden University},
	author = {Gomon, Georgy},
	year = {2022},
}

@article{rue_approximate_2009,
	title = {Approximate {Bayesian} {Inference} for {Latent} {Gaussian} models by using {Integrated} {Nested} {Laplace} {Approximations}},
	volume = {71},
	issn = {1369-7412},
	url = {https://doi.org/10.1111/j.1467-9868.2008.00700.x},
	doi = {10.1111/j.1467-9868.2008.00700.x},
	number = {2},
	urldate = {2025-09-26},
	journal = {Journal of the Royal Statistical Society Series B: Statistical Methodology},
	author = {Rue, Håvard and Martino, Sara and Chopin, Nicolas},
	month = apr,
	year = {2009},
	pages = {319--392},
}

@article{hastings_monte_1970,
	title = {Monte {Carlo} sampling methods using {Markov} chains and their applications},
	volume = {57},
	issn = {0006-3444},
	url = {https://doi.org/10.1093/biomet/57.1.97},
	doi = {10.1093/biomet/57.1.97},
	number = {1},
	urldate = {2025-09-26},
	journal = {Biometrika},
	author = {Hastings, W. K.},
	month = apr,
	year = {1970},
	pages = {97--109},
}

@misc{klyne_average_2025,
	title = {Average partial effect estimation using double machine learning},
	url = {http://arxiv.org/abs/2308.09207},
	doi = {10.48550/arXiv.2308.09207},
	urldate = {2025-09-26},
	publisher = {arXiv},
	author = {Klyne, Harvey and Shah, Rajen D.},
	month = jul,
	year = {2025},
	note = {arXiv:2308.09207 [math]},
}

@article{stoker_consistent_1986,
	title = {Consistent {Estimation} of {Scaled} {Coefficients}},
	volume = {54},
	issn = {0012-9682},
	url = {https://www.jstor.org/stable/1914309},
	doi = {10.2307/1914309},
	number = {6},
	urldate = {2025-09-26},
	journal = {Econometrica},
	author = {Stoker, Thomas M.},
	year = {1986},
	note = {Publisher: [Wiley, Econometric Society]},
	pages = {1461--1481},
}

@book{gelman_bayesian_1995,
	address = {New York},
	title = {Bayesian {Data} {Analysis}},
	isbn = {978-0-429-25841-1},
	publisher = {Chapman and Hall/CRC},
	author = {Gelman, Andrew and Carlin, John B. and Stern, Hal S. and Rubin, Donald B.},
	month = jun,
	year = {1995},
	doi = {10.1201/9780429258411},
}

\end{document}